\documentclass[twocolumn,showpacs,preprintnumbers,amsmath,amssymb]{revtex4}
\usepackage{graphicx}% Include figure files
\usepackage{dcolumn}% Align table columns on decimal point
\usepackage{bm}% bold math

\begin{document}

\title{
Spontaneous Collapse of Unstable Quantum Superposition State
}

\author{Takuya Okabe}
 \email{ttokabe@ipc.shizuoka.ac.jp}
\affiliation{
Faculty of Engineering, Shizuoka University, 
Hamamatsu 432-8561, Japan
}

\date{\today}

\begin{abstract}
On the basis of a 
proposed model 
of wave function collapse, we investigate 
spontaneous localization of a quantum state.
The model is similar to the Ghirardi-Rimini-Weber model,
while we postulate the localization functions to
depend on the quantum state to suffer collapse.
According to the model, 
dual dynamics in quantum mechanics, 
deterministic and stochastic time evolution,
are algorithmically implemented in tandem.
After discussing the physical implications of the model qualitatively,
we present numerical results for one-dimensional systems by way of
 example.
\end{abstract}

%03.65.Ta 	Foundations of quantum mechanics; measurement theory
%02.50.Ey 	Stochastic processes 
%07.05.Tp 	Computer modeling and simulation


\maketitle

\section{\label{Introduction}Introduction}

There has been a significant increase of interest 
in the foundations of quantum mechanics (QM).
This owes undoubtedly to the technological progress 
achieved during the last decades
in experimental investigation of 
how quantum systems behave
not only statistically, but on an individual level.
In fact, there are experimental results for microscopic systems
which appear to be naturally explained by resorting to 
collapse of wave function\cite{cd86,pp87,
sbnt86,co88,bz88,pk88,am91,rb88,
bhiw86,nsd86,lbmw03}.
Notwithstanding, 
wave function collapse has ever been the source of the problem
in understanding and interpreting QM,
and there is still no definitive consensus 
on the ``measurement'' problem.
Accordingly,
along with those which accept the collapse postulate,
there are many other interpretations,
such as
%accounting for apparent inconsistency.
de Broglie-Bohm causal interpretation\cite{bo52a,hol,dgz},
statistical interpretation\cite{b70},
decoherence\cite{ze70,mn80a,jz85,np93,gjkksz96,zu03},
modal interpretations\cite{vF91,he89,vd98},
consistent (decoherent) histories\cite{gr84,om92,gh90},
many-worlds/minds interpretations\cite{ev57,dg73,de85,al,a92}.
In this paper, we aim to investigate a realistic model 
of wave function collapse.

First of all, 
it is totally unsatisfactory to connect wave function collapse 
with an act of measurement,
because the concept of measurement 
is ill-defined\cite{bell87qmc,be90}.
To describe wave function collapse from a realistic standpoint,
we have to specify a well-defined dynamics of 
the stochastic process of collapse.
In this regard,
there are long-standing studies on
stochastic nonlinear equations to 
realize collapse effectively.
On the one hand, models to cause
rapid but {continuous} collapse
in the sense of Brownian motion
 were studied 
by Bohm and Bub\cite{bb66}, 
Pearle\cite{pe76},
Di\'osi\cite{di85}, and Gisin\cite{gi84a}.
On the other hand, a model postulating 
{discontinuous} instantaneous collapse was elaborated
in the pioneering work 
by Ghirardi, Rimini and Weber (GRW)\cite{grw86,bell87qj}.
At present, the GRW model and the Pearle model have been jointly 
developed to the 
continuous spontaneous localization (CSL) model\cite{gpr90}.
Mathematically, CSL
is based on 
a stochastically modified Schr\"odinger equation.
The current status of studies 
on what are and have to be achieved in their models
has been reviewed extensively\cite{gh00,bg03,p99,prca99}.
On the basis of 
the idea of spontaneous collapse (SC),
which may have a clue to the measurement problem,
a good deal of related works 
are published indeed\cite{fr90,mi91,pe95,hu96,ah00}.
The task of theories based on and accounting for SC
is to specify when, how and how often the collapse occurs,
and to investigate the physical consequences.
This is particularly important because
the effect of SC must be physically relevant,
in striking contrast to the other interpretations.
The predictions of collapse theories 
should more or less deviate from
those of standard quantum mechanics (SQM).

In this paper, we propose and investigate
another such model of similar physical implications
as the original GRW model.
We also introduce two constants to characterize the dynamics of
collapse.
In the GRW model, 
a wave function is subjected to incessant spontaneous localizations,
or the wave function is multiplied by a localization function,
e.g., the Gaussian function,
of which the frequency and the localization width
are postulated to be universal constants, 
$\lambda \simeq 10^{-16}$sec$^{-1}$ and $a\simeq 10^{-5}$cm,
respectively.
The GRW localization mechanism is such that 
its frequency 
increases as the number $N$ of constituents of 
a composite system increases,
so that a macroscopic object comprising
an Avogadro number of constituents 
collapses extremely rapidly,
at a rate of
$N\lambda \simeq 10^7$sec$^{-1}$.
In contrast, we propose to postulate 
a constant collapse rate
$\gamma_0=\tau_0^{-1}$,
while we model that the localization functions,
particularly its length scale, are variable and 
depend on the wave function to suffer collapse.
In effect, we introduce an energy scale $T_0$
so that the stochastic collapse should occur
to the effect to cost energy $\Delta E\simeq T_0$.
Therefore, in our model, 
the macroscopic number $N$ does not play any essential role,
so that the effect of collapse may manifest itself 
even in microscopic systems\cite{toappear}.

After introducing our collapse model 
in Section \ref{sec:2-1}, 
we discuss the physical implications of the model 
in Section \ref{sec:2-2} and in Section \ref{sec:3}.
By way of illustration, in Section \ref{sec:4},
we apply the modified dynamics to obtain numerical results for
one-dimensional systems.
Although
we are mainly interested in the individual dynamical evolution of a
particular system,
we briefly discuss
the statistical description of the system
in terms of the density matrix
in Section \ref{sec:discussions}.
We state our conclusion in Section \ref{sec:conclusion}.
In Appendix \ref{ap:vari}, 
we give an analytical treatment of a simple special case
of the model
to support general discussions in the main text.
We discuss a generalization required
to treat many particle systems in Appendix \ref{ap:many}.
The main purpose of this paper is to show that the proposed model
provides us with a consistent picture of crossover between quantum and
classical mechanics.

\section{Model}\label{sec:2}
\subsection{Single particle problem}\label{sec:2-1}

\begin{figure}[t]
\begin{center}
\includegraphics[width=0.5\textwidth]{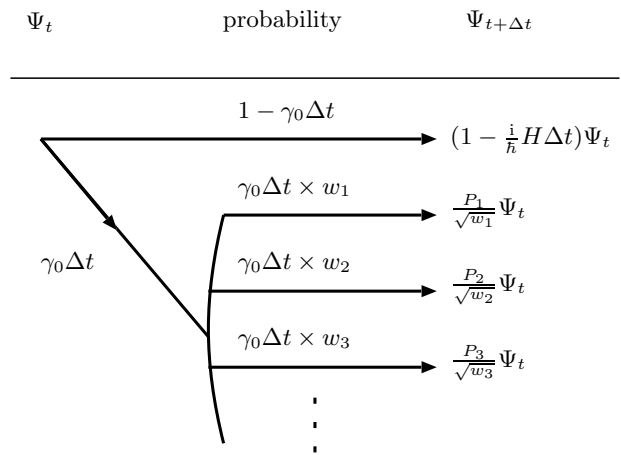}
\caption{
Flow chart of the modified dynamics.
From the initial state in the left,
one (and only one) of the states in the right column is 
realized with the probability attached to the arrow.
}\label{fig:chart}
\end{center}
\end{figure}

We discuss a non-relativistic model of spontaneous collapse.
Let us consider the normalized state
$\Psi_{t}({\bf r})$ as a function of space ${\bf r}$ and time $t$,
where the indices representing
internal quantum degrees of freedom
like spins are suppressed for simplicity.
We postulate that
$\Psi_{t+\Delta t}({\bf r})$ 
after the infinitesimal elapse of time $\Delta t$
is 
determined from 
$\Psi_{t}({\bf r})$ 
stochastically 
as follows:
(i) 
With the probability $\gamma_0 \Delta t$,
\begin{equation}
 \Psi_{t+\Delta t}({\bf r})
=
\left(
1-\frac{{\rm i}}{\hbar}\hat{H}\Delta t
\right)
 \Psi_{t}({\bf r}),
\label{1stlaw}
\end{equation}
while (ii)
\begin{equation}
 \Psi_{t+\Delta t}({\bf r})
=
\Phi_n({\bf r})\equiv
\frac{1}{\sqrt{w_n}}
P_n({\bf r}) \Psi_t({\bf r}),
\label{2ndlaw}
\end{equation}
with the probability $(1-\gamma_0)w_n \Delta t$, 
where
\begin{equation}
 w_n\equiv 
%\langle \Phi_n|\Phi_n\rangle=
\int{\rm d}{\bf r} 
|\Psi_t({\bf r})|^2 P_n({\bf r})^2.  
\label{wndef}
\end{equation}
In (\ref{wndef}),
the sum over the suppressed indices is implicit
in the definition of the modulus square 
$\rho({\bf r})=|\Psi_t({\bf r})|^2$,
which thus represents the density 
in ordinary space.
Unlike (i), 
the process (ii) is discontinuous by postulate,
so that it cannot be put into a differential equation,
though we can take the limit $\Delta t\rightarrow 0$
without affecting the physical content of the model.
The branching structure of the proposed dynamics
is schematically summarized in Fig.~\ref{fig:chart}.
According to the modified dynamics,
the quantum state $\Psi({\bf r})$ is subjected to 
collapse (ii) with the constant rate $\gamma_0$.
The probability to keep following the deterministic evolution (i)
for $N\Delta t=\tau$ is given by
%for $N$ times $\Delta t=\tau/N$ is given by
\(
\left(1-\frac{\gamma_0 \tau}{N}\right)^N \rightarrow 
\exp\left(-\gamma_0 \tau
\right)
\) as $N\rightarrow \infty$.
The collapse rate $\gamma_0$, or the time $\tau_0=\gamma_0^{-1}$, 
is the first parameter of our model.

In (\ref{2ndlaw}), the effect of collapse is described 
by the real-valued continuous envelop functions $P_n({\bf r})$,
which we call the localization functions, 
following the GRW model.
%which correspond to the localization function of the GRW model.
%to multiply the initial state $\Psi_t({\bf r})$.
The factor $\sqrt{w_n}$ in (\ref{2ndlaw}) is 
to normalize
%for the sake of the normalization of 
the collapse outcomes, $\langle
\Phi_n|\Phi_n\rangle=1$.
The functions $P_n({\bf r})$
%
% $P_n({\bf r})$ are real  
%in (\ref{2ndlaw}) and (\ref{wndef}),
%which depend on the initial state $\Psi_t({\bf r})$,
have to satisfy
\begin{equation}
 \sum_n P_n({\bf r})^2=1,\qquad  0\le P_n({\bf r})\le 1,
\label{sumnPnP2=1}
\end{equation}
in order to meet the conservation of probability,
%for $w_n$ defined in (\ref{wndef}) to meet
\begin{equation}
 \sum_n w_n =1.
\end{equation}
%
%to be regarded as probabilities,
%which must sum up to unity,
%multiply $\Psi_t({\bf r})$ with the envelop function $P_n({\bf r})$.

 %supports of $P_n({\bf r})$ may overlap generally.

To achieve spontaneous localization,
the GRW model assumes
$P_n({\bf r})$
to be the Gaussian functions with a fixed width
centered at random positions.
We postulate rather that a set of the real functions $P_n({\bf r})$ 
is determined optimally
depending on the state $\Psi_t({\bf r})$
so as to maximize
\begin{equation}
 S'= \sum_n w_n \int |\Phi_n|^2
\log |\Phi_n|^2 {\rm d}{\bf r},
%--  S'= \sum_n w_n \int |\Phi_n|^2(\log |\Phi_n|^2 -1 ) {\rm d}{\bf r},
\label{S'0}
\end{equation}
under the constraint
\begin{equation}
T_0 \Delta S = \Delta E,
\label{T0DeltaS=DeltaE}
\end{equation}
where 
\begin{eqnarray}
 \Delta S&=&-
{\rm Tr}(\hat{\rho}\log \hat{\rho})
%\sum_n w_n \log w_n,
\label{DeltaS}
\end{eqnarray}
and
\begin{eqnarray}
 \Delta E&=&
{\rm Tr}(\hat{\rho} \hat{H})
%\sum_n w_n \langle \Phi_n| \hat{H}|\Phi_n\rangle
-
\langle \Psi| \hat{H}|\Psi\rangle.
\label{DeltaE}
\end{eqnarray}
are the physical changes of entropy and energy due to collapse.
They are defined customarily
in terms of the density matrix 
$\hat{\rho}=\hat{\rho}_{t+\Delta t}$
of the collapse outcome.
%to be obtained after the collapse.
%
%define the change of entropy and energy
%caused by the postulated collapse (\ref{2ndlaw}),
%
In real space representation, 
the matrix elements of $\hat{\rho}$ are given by
\begin{eqnarray}
 \langle {\bf r}|\hat{\rho}|{\bf r'}\rangle
&=&\sum_n w_n \Phi_n({\bf r})^*\Phi_n({\bf r}').
%\nonumber\\
%&=&\Psi({\bf r})^*\Psi({\bf r}')\sum_n P_n({\bf r})P_n({\bf r}').
\label{densitymatrix} 
\end{eqnarray}

The above is the procedure that we propose to fix 
the collapse outcomes $\tilde{\Phi}_n=P_n{\Psi}$
resulting from an arbitrary state $\Psi$.
From our standpoint to regard wave function collapse 
as a real process,
(\ref{DeltaS}) and (\ref{DeltaE}) represent 
the physical changes accompanied by the collapse,
both of which are the important quantities to 
characterize the stochastic irreversible process.
Thus (\ref{T0DeltaS=DeltaE}) and 
(\ref{sumnPnP2=1}) respectively
represent statistical conservation of `free energy' $F=E-T_0S$
and the density,
%\begin{equation}
%|\Psi({\bf r})|^2
%= 
%\sum_n w_n |\Phi_n({\bf r})|^2.
%\end{equation}
\begin{equation}
|\Psi({\bf r})|^2
= 
\sum_n |\tilde{\Phi}_n({\bf r})|^2.
\end{equation}
%where $\tilde{\Phi}_n=P_n{\Psi}$ and
%$w_n=\int{\rm d}{\bf r}|\tilde{\Phi}_n({\bf r})|^2$.
Under these statistical constraints,
we obtain
the set of collapse outcomes $\tilde{\Phi}_n$ 
by modifying to replace the squared amplitude $\rho$ of 
the initial state
$\Psi=\sqrt{\rho} {\rm e}^{{\rm i}\alpha}$
with $\tilde{\rho}_n\equiv P_n^2 \rho$ 
without affecting the phase $\alpha$.
Thus, in a sense, 
we envisage the collapse as a kind of 
fragmentation of wave packet $\rho\rightarrow \{\tilde{\rho}_n\}$
in real space.

In case where there happens to be non-equivalent sets of 
the localization functions $\{P_n\}$
for given $\Psi_t({\bf r})$,
%there happen to be 
%equivalent sets $\{P_n\}$ for given $\Psi_t({\bf r})$ and 
%and $(\delta {\bf r})^2$,
e.g., by reason of symmetry, 
we require another postulate that a single set 
is chosen from among them
at random with equal a priori probability.
It would be clear that these postulates 
to formulate the probabilistic `second law' (\ref{2ndlaw})
are adopted
on the analogy drawn from Thermodynamics of 
the irreversible process of open systems 
immersed in a heat bath at temperature $T_0$.
The constant $T_0$ is the second parameter of our model.

We adopt (\ref{S'0}) as a proper expression to 
implement the spontaneous tendency of collapse
to reduce the spatial extent of the quantum state,
\begin{eqnarray}
(\delta {\bf r})^2&=&
\sum_n w_n 
\left(
{\langle \Phi_n| {\bf r}\cdot {\bf r}|\Phi_n\rangle} 
-|{\langle \Phi_n| {\bf r}|\Phi_n\rangle}|^2 
\right).
\label{delr^2}
\end{eqnarray}
Thus we take account of 
the special role of the position basis in terms of $S'$. 
In effect, 
to minimize the variance $(\delta {\bf r})^2$ means to maximize 
the second term in (\ref{delr^2}),
\begin{equation}
\sum_n w_n 
|{\langle \Phi_n| {\bf r}|\Phi_n\rangle}|^2 
=\sum_n w_n 
\left|
\int 
|\Phi_n ({\bf r})|^2
{\bf r}  {\rm d}{\bf r}
\right|^2,
\label{sumnwn|int|2} 
\end{equation}
for the first term is independent of $\Phi_n$,
\[
\sum_n w_n 
{\langle \Phi_n| {\bf r}\cdot {\bf r}|\Phi_n\rangle} 
= \langle \Psi| {\bf r}\cdot {\bf r}|\Psi\rangle.
\]
The expression in (\ref{sumnwn|int|2}) is not easy to treat with
formally.
To serve our purpose of localization equally well,
we shall rather adopt the expression (\ref{S'0}).
The positive sign in (\ref{S'0})
is opposite to the conventional definition of entropy
with a minus sign 
%in terms of a distribution function $\rho({\bf r},{\bf p})$
(cf. (\ref{DeltaS})).
This is because
we intend to describe the spontaneous tendency 
toward compact localization (instead of diffusion)
of quantum states
in terms of the physical quantity $S'\simeq \log (\delta {\bf r})^{-3}$
to be maximized.
%
%that 
%compactly localized states are prefered to extended ones.
%
%
%In terms of $S'$, we can describe mathematically the tendency that 
%compactly localized states are prefered to extended ones.
%in six dimensional phase space,
%to replace $|\Phi_n({\bf r})|^2$.
% in three dimensional space.
%As far as we treat only single-particle problems, 
For simplicity, for the same purpose,
we might as well maximize a simpler expression like
\begin{equation}
\overline{\rho^2}
=\sum_n w_n \int 
{\rm d}{\bf r}|\Phi_n ({\bf r})|^4.
\label{rho2}
\end{equation}
Nevertheless, 
(\ref{S'0}) has a physically preferable feature
for the purpose of generalization (See Appendix~\ref{ap:many}).

%, nevertheless,
%(\ref{S'0}) is physically preferable 
% preferable expression

%
%it is mathematically favorable 
%to introduce
%the more concise expression than $(\delta {\bf r})^2$ in (\ref{delr^2}) 
%to figure out the localization functions $P_n$.

%instead of (\ref{delr^2})

It is stressed that 
the physical implications of the proposed modified dynamics,
which are discussed in what follows,
 are not affected qualitatively 
by whichever quantity one shall adopt to realize localization.
%
%by a specific choice of the quantitity to maximize.
%
The results of our primary concern are essentially 
the consequences of the very idea that 
stochastic collapse tends to reduce the spatial extent
of wave packet as small as possible
to the extent to cost energy $\Delta E\simeq T_0$ 
by (\ref{T0DeltaS=DeltaE}),
or 
that the collapse outcomes are variably determined 
depending on the state $\Psi$ to be reduced.

%so determined variably 
%the collapse outcomes
%depending on the state $\Psi$ to be reduced.

\subsection{Quantum Limits}\label{sec:2-2}

By construction, 
the non-trivial effect of collapse becomes negligible 
in the limit $\gamma_0\rightarrow 0$.
Consequently, in the time regime appreciably shorter than $\tau_0$, 
the quantum results based on the deterministic evolution (i)
remain intact.
%are reproduced without modification.
Hence, we may regard
the short time regime $t \ll \tau_0$ as a quantum regime.
%In effect, the parameter $\tau_0$ is 

%and 
%the long time regime $t \gg \tau_0$ as classical.

In the long run $t \gg \tau_0$, however,
the standard quantum predictions 
based on the strict validity of the unitary evolution (i) alone
have to be modified more or less.
One of the most notable consequences of the collapse model
is the violation of energy conservation $\Delta E\ne 0$\cite{grw86,gpr90},
which is taken for granted in the above model through 
(\ref{T0DeltaS=DeltaE}).
%In effect, in the above scheme,
%conserved is the `free energy' $F=E-T_0S$ instead, 
%the energy production $\Delta E$ due to collapse 
%is regulated by the constraint (\ref{T0DeltaS=DeltaE}).
%the spontaneous irreversible collapse (\ref{2ndlaw}) 
%violates the energy conservation law.
With the physical entropy $\Delta S$ created upon collapse,
%by the collapse of the initially pure ensemble of $\Psi$,
the rate of energy production is given by $T_0\Delta S/\tau_0$,
which must be negligibly small in practice.

For definiteness, let us discuss the system described by 
the Hamiltonian
\begin{equation}
\hat{H}=-\frac{\hbar^2 \nabla^2}{2m} +V({\bf r}),
\label{model} 
\end{equation}
for which we obtain
\begin{eqnarray}
 \Delta E&=&
\frac{\hbar^2}{2m}
\int {\rm d}{\bf  r} \rho({\bf r})
\sum_n 
\left(\nabla P_n({\bf r})
\right)^2 \ge 0,
\label{DeltaE2}
\end{eqnarray}
without approximation.
It is noted that 
(\ref{DeltaE2}) depends on the potential $V({\bf r})$
only implicitly through the `probability' density
\[
\rho({\bf r})=|\Psi_t({\bf r})|^2.
\]
In general, for an observable $\hat{O}$, 
we obtain
\begin{eqnarray}
 \Delta O
&=&\sum_n w_n 
\langle \Phi_n| \hat{O}|\Phi_n\rangle
-
\langle \Psi| \hat{O}|\Psi\rangle
\nonumber\\
%&=&
%\int{\rm d}{\bf r} 
%\Psi_t({\bf r})^* 
%\left(
%\sum_n 
%P_n({\bf r}) \hat{O}
%P_n({\bf r})
%-\hat{O}
%\right)
%\Psi_t({\bf r}) 
%\nonumber\\
&=&
\int{\rm d}{\bf r} 
\Psi_t({\bf r})^* 
\sum_n
P_n({\bf r})
[\hat{O},P_n({\bf r})]_-
\Psi_t({\bf r}) 
%\langle \Psi|  [\hat{O},P_n({\bf r})]_-|\Psi\rangle.
\end{eqnarray}
Even when $\hat{O}$ and $P_n({\bf r})$ commute,
\(
 [\hat{O},P_n({\bf r})]_- =0,
\)
the collapse (\ref{2ndlaw}) conserves
the expectation value $\langle \hat{O}\rangle$
only statistically, $\Delta O=0$, 
so that we should generally expect 
$\langle \Phi_n| \hat{O}|\Phi_n\rangle
\ne
\langle \Psi| \hat{O}|\Psi\rangle
$
individually.
%Keeping this in mind, 
%let us consider 
For instance, 
the center of mass $\langle {\bf r}\rangle$ 
%for $\hat{O}={\bf r}$ 
will fluctuate
as we follow the individual behavior of 
a particular state evolving
according to the modified dynamics
(cf. Fig.~\ref{fig:free2}).

According to (\ref{T0DeltaS=DeltaE}) and
(\ref{DeltaE2}),
in the limit $T_0\rightarrow 0$,
we conclude $P_n({\bf r})=$ constant 
if $\rho({\bf r})>0$ for any ${\bf r}$.
%for the states
%with the density $\rho({\bf r})$ that does not vanish all over the space, 
%viz., for $\rho({\bf r})>0$.
%
%In the the limit $T_0\rightarrow 0$,
%where we must conclude $P_n({\bf r})$ to be constant
%according to (\ref{DeltaE2}), 
%if we have $\rho({\bf r})>0$ all over the space.
%Therefore, 
Therefore, in this case, 
we recover not only the energy conservation $\Delta E=0$,
but also the quantum limit with no collapse effectively.
This is because
the collapse (\ref{2ndlaw}) with
the constant $P_n$
%, independent of ${\bf r}$,
has no physical effect at all, for
the collapse outcomes 
become physically all equivalent to 
the pre-collapse state,
$\tilde{\Phi}_n({\bf r})= P_n\Psi \propto \Psi({\bf r})$.
Hence, as well as $\gamma_0\rightarrow 0$,
the limit $T_0\rightarrow 0$
may also be called the quantum limit.
Nevertheless, in contrast to the former limit,
all the standard quantum results are 
not recovered in the latter.
As discussed below in Section \ref{sec:3-3},
our modified dynamics eventually gives predictions 
against SQM even at $T_0=0$.

\section{Qualitative Results}\label{sec:3}

\subsection{Classical Limit}\label{sec:3-1}

According to the modified dynamics presented above, 
the spatial extent of the wave packet $\Psi({\bf r})$
is regularly truncated down to a finite size.
Without detailed calculation,
we can make an order estimate of the length scale of 
the wave packet under spontaneous collapse.
% of the size of wave packet.
%how small it is reduced.  $(\delta {\bf r})^2$ 
Owing to the constraint (\ref{T0DeltaS=DeltaE}),
even in free space $V({\bf r})=0$,
the energy scale $T_0$ introduces the length 
\begin{equation}
\lambda_0\equiv 
\hbar\sqrt{\frac{2\pi}{mT_0}},
\label{lam0}
\end{equation}
which is nothing but 
the thermal de Broglie wavelength at ``temperature'' $T_0$.
Accordingly, we conclude the finite wave packet width of order
$|\delta {\bf r}| \simeq \lambda_0 \propto m^{-1/2}$,
which brings about our purpose to reproduce classical mechanics 
of the point mass ${\bf r}(t)=\langle {\bf r}\rangle$
in the macroscopic limit $m\rightarrow \infty$ where
$\lambda_0\rightarrow 0$,
%appropriate for macroscopic bodies,
without introducing any external observer.
The finite width (\ref{lam0}) from (\ref{T0DeltaS=DeltaE}) 
is just as expected 
by the thermodynamic analogy
mentioned above.
In effect, the degree of localization $(\delta {\bf r})^2$
depends on
the energy $\Delta E>0$ required for the localization.
If it were not for the constraint (\ref{T0DeltaS=DeltaE}),
or if we let $T_0\rightarrow \infty$,
% to replace (\ref{T0DeltaS=DeltaE}) with $\Delta S>0$,
the hypothetical spontaneous collapse would reduce
$(\delta {\bf r})^2$ without limit,
and the established results 
of quantum mechanics at a short length scale 
must be spoiled altogether.
Thus, in the present model, 
the crossover length scale presumed between 
classical and quantum regimes
is brought in by the non-trivial scale $T_0$ in a controlled manner.

The position $\langle {\bf r}\rangle$ of the free particle
fluctuates with 
the amplitude of the order of $|\delta {\bf r}|\simeq \lambda_0$. 
In principle, the finite value of $|\delta {\bf r}|$ 
%predicted for a macroscopic body
should 
give rise to 
non-trivial quantum corrections to 
Newton's equation of classical mechanics
for the approximately well-defined position ${\bf r}(t)$ of the body.
Nevertheless, 
it has to be remarked that the non-trivial effects due to 
$|\delta {\bf r}|\simeq \lambda_0$ can not necessarily be 
conspicuous in real situations,
because they may be completely masked
by environmental decoherence,
the effects of which have been intensively discussed for
decades\cite{te93,tvg95,gjkksz96,zu03}.
In effect, as we see below,
the scale $T_0$ of collapse
should be substantially subtler than any perturbations of practical
relevance.
Still it would not be impractical to expect that the non-trivial
predictions of the model, namely,
the intrinsic decoherence of the constant rate $\gamma_0$,
might survive the extrinsic disturbances in some controlled situations.

\subsection{Born's rule}\label{sec:3-2}

%for which $\lambda_0$ sets the smallest length scale,
%$m\rightarrow \infty$,
Let us turn our attention to
the microscopic system for which 
$\lambda_0$ defined in (\ref{lam0})
is substantially larger than 
the typical length scale 
$|\delta {\bf r}|\simeq \sigma$
of the system.
The latter is
determined by the kinetic as well as potential terms in
the Hamiltonian.

To begin with, let us consider 
the wave packet 
%density $\rho({\bf r})$
whose linear dimension is of order $\sigma$.
We represent by $\lambda$
the typical length scale over which 
the localization functions $P_n$ change appreciably.
As we noted at the end of the last section,
if $\rho({\bf r})>0$ for any ${\bf r}$, then 
we should obtain $\lambda\simeq \lambda_0 \gg \sigma$.
%
%for the typical length scale $\lambda$ over which 
%the localization functions $P_n$ change appreciably.
%
%
As a result,
we conclude that the collapse has little effect on the quantum state,
viz., $P_n\Psi\propto \Psi$.
%
%%, which does not vanish anywhere,
%If $\sigma$ is larger than 
%the typical length scale $\lambda$ over which 
%the localization functions $P_n$ change appreciably,
%then the condition 
%%$\Delta E= T_0\Delta S$ 
%(\ref{T0DeltaS=DeltaE})
%gives 
%$\lambda\simeq \lambda_0 \gg \sigma$, 
%so that we conclude that $P_n\Psi\propto \Psi$,
%or 
%the localization functions $P_n$
%does no physically affect the state $\Psi$ under consideration,
%because $P_n$ then may be essentially regarded as constant.
For example,
in the limit $\sigma\ll \lambda_0$,
the localized wave packet 
\begin{equation}
\phi_\sigma({\bf r})=
\frac{1}{(2\pi \sigma^2)^{3/2}}
{{\rm e}^{-|{\bf r}|^2/2\sigma^2}}
\label{gaussian}
\end{equation}
is hardly affected by the postulated collapse.
Hence the effect of (ii) may be 
schematically expressed as
\begin{equation}
\phi_\sigma({\bf r}) \rightarrow \phi_\sigma({\bf r}).
\label{phisigtophisig}  
\end{equation}
We conclude $\Delta E=T_0 \Delta S\simeq 0$ accordingly.
%The microscopic states like (\ref{gaussian})
%are not affected by the modified dynamics.
%

Next, we consider the case in which 
the opposite limit $\lambda\ll \sigma$ may be realized,
where the postulated collapse has a drastic effect.
Since $\sum_n (\nabla P_n({\bf r}))^2$ in (\ref{DeltaE2})
takes a finite value $\sim \lambda^{-2}$ 
for ${\bf r}$ in the domain 
where $P_n$ varies,
% over a linear width $\sim \lambda$,
%vanishes 
%except in the region of the width $\sim \lambda$
%at the boundaries where
%the localization functions $P_n({\bf r})$ decay from 1 to 0,
%
we obtain
\begin{equation}
\Delta E\simeq 
\frac{\hbar^2}{2m} 
\frac{q}{\lambda^2},
\label{Delta Esimeqhbar2} 
\end{equation}
in which the factor
\(q=
\int_\lambda \rho({\bf r}){\rm d}{\bf r}
\)
denotes the spatial integral over the `boundary' of ${P_n({\bf
r})}$ where
 $P_i({\bf r})P_j({\bf r})\ne 0$ for $i\ne j$ (see (\ref{varep=q})).
Hence, by 
the constraint $\Delta E=T_0 \Delta S\simeq T_0$, we obtain
\begin{equation}
 \lambda\simeq \lambda_0 \sqrt{q},
\label{lambdasimeqlambda0q} 
\end{equation}
or
\begin{equation}
 \lambda\simeq \lambda_0^2\bar{\rho}, 
\label{lambdasimeq}
\end{equation}
in terms of the density per length $\bar{\rho}$ defined by $q=\lambda
\bar{\rho}$. 
Therefore, $\lambda$ can 
%become small 
%substantially smaller than $\sigma$
%
be scaled down by the factor $q$ or $\bar{\rho}$, 
which may happen to be exponentially small, e.g.,
when $\rho({\bf r})$ is represented by 
a linear superposition of well separated wave packets.

For example, let us consider a typical case of 
an `unstable' linear superposition 
of the localized wave packet (\ref{gaussian}),
\begin{equation}
 \Psi({\bf r})=c_+ \phi_\sigma({\bf r}-{\bf R})
+c_- \phi_\sigma({\bf r}+{\bf R}),
\label{psi=c+phi+c-}
\end{equation}
where $|c_+|^2+|c_-|^2=1$.
For $\Psi({\bf r})$ with $|{\bf R}|\gg \sigma$, 
we find $P_\pm({\bf r})$ to give
%with $\lambda\ll \sigma$
$P_\pm\Psi\simeq c_\pm \phi_\sigma({\bf r}\mp{\bf R})$.
As a result of collapse, therefore, we obtain
\begin{equation}
% \Psi\rightarrow \frac{c_\pm}{\sqrt{w_\pm}}\phi_\sigma({\bf r}\mp {\bf R})
 \Psi\rightarrow \frac{c_+}{\sqrt{w_+}}\phi_\sigma({\bf r}- {\bf R}),
\quad {\rm or}\quad
\frac{c_-}{\sqrt{w_-}}\phi_\sigma({\bf r}+ {\bf R})
\label{Psitophisigma} 
\end{equation}
with the probabilities $w_{+}=|c_+|^2$ and
$w_{-}=|c_-|^2$, respectively.
%$w_{\pm}=|c_\pm|^2$.
The point is that the two states
$\phi_\sigma({\bf r}+ {\bf R})$ and $\phi_\sigma({\bf r}-{\bf R})$
have little spatial overlap, 
so that it hardly costs energy to 
reduce 
$|\delta {\bf r}|\simeq 2 ({w_+w_-})^{1/2}
|{\bf R}|\gg \sigma
$ of the initial wave packet (\ref{psi=c+phi+c-})
%$\Psi$
down to 
%a minimum size
$|\delta {\bf r}|\simeq \sigma$ for 
(\ref{Psitophisigma}),
which is `stable' in the sense of (\ref{phisigtophisig}).
Indeed, for definiteness, owing to 
$\bar{\rho}\propto {\rm e}^{-|{\bf R}|^2/\sigma^2}/\sigma $,
we obtain $\lambda \ll \sigma$ for $|{\bf R}|\gg \sigma$.
Thus the localization functions 
are approximately represented by
the step functions 
$P_-({\bf r})\simeq \theta({\bf r}\cdot {\bf R})$
and
$P_+({\bf r})\simeq \theta(-{\bf r}\cdot {\bf R})$,
where $\theta(x)=1$ for $x>0$ and $\theta(x)=0$ for $x<0$
(cf. Figs.~\ref{fig:doublewell2} and \ref{fig:Pn}).
This is essentially 
how Born's rule follows from our modified dynamics.

The above argument 
is based on the emergence of the different scales 
$\lambda\ll \sigma\ll \lambda_0$
through (\ref{T0DeltaS=DeltaE}).
As a function of the separation $|{\bf R}|$,
the qualitative change of behavior from (\ref{phisigtophisig})  
to (\ref{Psitophisigma})
occurs around $|{\bf R}|\simeq \sigma$,
which is independent of $T_0$.
In other words, 
according to (\ref{lambdasimeqlambda0q}),
the microscopic scale $\lambda$ for $P_n$
may be derived as the product of $\lambda_0$,
which must be larger than any other length scales of
a microscopic quantum system,
times the exponentially small factor $\sqrt{q}$.
%due to the low density $\rho({\bf r})$ at the
%boundary of the collapse outcomes.
Although the former depends on $T_0$,
it is essentially the latter factor $q$ 
that provides us with the 
smallest scale $\lambda$ of $P_n$.
Thus the qualitative features discussed above do not 
depend sensitively on the precise value of $T_0$,
but are mainly determined by the density distribution
of the state $\Psi({\bf r})$.

%As in (\ref{lambdasimeqlambda0q})
%(\ref{lambdasimeq})

\subsection{Predictions against SQM}\label{sec:3-3}

\begin{figure}
\begin{center}
%/home/okabe/tex/ohp/doubleslit1b2.eps
%\includegraphics[width=0.4\textwidth]{doubleslit.eps}
\includegraphics[width=0.4\textwidth]{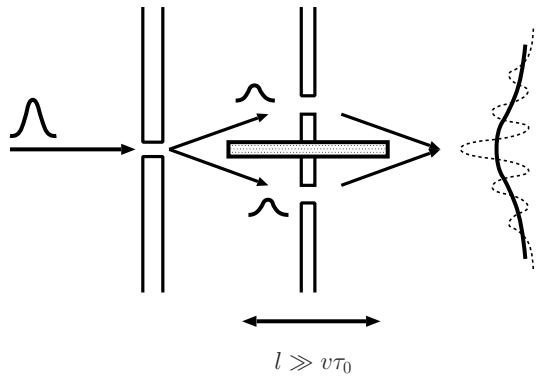}
\caption{
According to the modified dynamics,
%we introduce the time $\tau_0$ beyond which 
%a linear superposition of wave packets cannot survive intact.
%Accordingly, 
the interference fringe pattern 
due to split and recombined wave packets
is predicted to be washed out,
if 
the distinct wave packets in a superposition had been
spatially separated further than their own widths 
for a duration longer than 
the collapse time 
$\tau_0$ before the recombination,
e.g., 
if the length $l$ of the 
impenetrable 
partition slab separating the two paths is
made larger than a coherence length $\sim v \tau_0$, 
where $v$ is the velocity of the incoming wave packet.
%the quantum fringe pattern 
%due to the interference of the wave packets
%is predicted to be washed out
%according to the present modified dynamics.
}
\label{fig:ds}
\end{center}
\end{figure}
The modified dynamics
describes the `measurement' involving wave function collapse 
%following Born's probability rule 
as a stochastic process to 
minimize $|\delta {\bf r}|$ of the wave function $\Psi$.
It serves our purpose that those classically inadmissible
superposition states of well separated wave packets like (\ref{psi=c+phi+c-})
are destabilized to decay with the finite lifetime $\tau_0$.
In principle, 
the non-trivial physical effects of the model
should be manifested themselves
in the interference experiments which are
devised to distinguish
the pure state $\Psi$ from the statistical mixture of 
the collapse outcomes $\Phi_n$.
For instance, we are led to an unconventional prediction 
of the destruction of the quantum interference pattern,
as schematically presented in Fig.~\ref{fig:ds}.
Since such an intrinsic decoherence is concluded
independently of the actual distance between the separated
wave packets in superposition, 
we must also conclude
%that we are necessarily led to 
seemingly unfavorable consequences 
for the microscopic single-particle states with nodes
along which the density vanishes, $\rho({\bf r})=0$.
%
%microscopic systems
%of an isolated single particle 
%in a quantum state 
%described by a wave function with nodes
%%single-particle states with nodes
%%, namely, anti-bonding orbitals isolated in general,
%is affected by collapse.
%For example, i
In contrast to
(\ref{phisigtophisig}) for (\ref{gaussian}), e.g., 
the wave function
$\phi_\pi({\bf r})\propto x\phi_{\sigma}({\bf r})$
for $\sigma\ll \lambda_0$
cannot but be halved
to result in the effectively nodeless
collapse outcomes $P_\pm\phi_{\pi}$;
\[
\phi_{\pi}({\bf r})
\longrightarrow 
\theta(\pm x) 
\phi_{\pi}({\bf r}),
\]
as one obtains
\(
 \lambda\simeq \sigma^3/\lambda_0^2 \ll \sigma
\)
in this case.
In short, 
the modified dynamics predicts that
wave functions with nodes generally have to be nodeless eventually.
%\theta(-x) x\phi_{\sigma}({\bf r})
It is clear that such drastic unconventional 
collapse processes must not intervene 
the conventional Schr\"odinger time evolution too frequently.
Thus we have to require the collapse time $\tau_0$ to be substantially
slower than any practical coherence time scales 
%of conventional microscopic processes 
for the quantum interference of spatially separated wave packets.

%*************** 
%%for they erase the off-diagonal coherence between the different
%%outcomes which are supposed to sustain for ever. 
%%
%%to the extent that
%
%In order that the effects of the postulated collapse 
%are practically of little relevance 
%as far as the 
%the final statistical outcomes of `measurement'
%are concerned,
%
%In general,
%we have to require the collapse time scale $\tau_0$ to be 
%substantially slower than any 
%coherence time scales of conventional
%microscopic processes involved
%in the quantum interference phenomena.

\subsection{Numerical estimate of the model parameters}\label{sec:3-4}

In practice, 
the numerical value for $T_0$ must be properly chosen such that
the characteristic length $\lambda_0$ for microscopic particles like
electron and neutron must be too large while 
the intrinsic wave packet widths
$\lambda_0$ for macroscopic bodies, 
which may be effectively regarded as material points,
are too small 
to be unexpectedly detected experimentally.
%(either too large or small for the crossover regime
%to be detected unexpectedly),
%
To put it concretely, in order to guarantee 
$\lambda_0 > 10$m for electron
while $\lambda_0 < 1\mu$m for a tiny particle of a nanogram,
we estimate
\(
10^{-49}\ {\rm J} <T_0< 10^{-39}\ {\rm J}. 
\)
Hence there is a relatively wide range open for $T_0$,
owing to the above note that
the model is relatively insensitive to $T_0$.

On the other hand, 
we must assume 
the collapse time $\tau_0$ to be
larger than a typical coherence time scale of 
microscopic quantum phenomena, as mentioned 
in the last subsection.
Nonetheless, at the same time, $\tau_0$ must be shorter than 
a typical scale of 
macroscopic classical phenomena
in order to warrant definite outcomes for 
classical bodies approximately at any time, i.e., 
to prevent such an embarrassing superposition state
of `classical' states with `distinct' spatial configurations
from developing and lasting for a long while.
%arbitrarily long time.
Therefore one may make a rough estimate like
$10^{-6} {\rm sec} \ll \tau_0 \ll 10^{-1} {\rm sec}$,
so we find that
the parameter range allowed for $\tau_0$
is more restricted than $T_0$.
In fact, we consider that
whether we could assume $\tau_0$ properly
would be the crucial point 
for the present model to be viable physically.
In what follows, 
we discuss the implications of the model
assuming that
experiments do not exclude room for the model parameters.
%$\tau_0$.

Without being affected by the collapse,
the wave packet evolves according to (\ref{1stlaw})
for the duration of order $\tau_0$.
In the meantime, 
the width $(\delta {\bf r})^2\simeq \lambda_0^2$
for a free particle
grows up to $\lambda_0^2\left(1+ 
(\frac{T_0 \tau_0}{4\pi \hbar})^2)\right)$,
so that the second length scale must be introduced if
$T_0\tau_0\gg 4\pi \hbar$.
However, with the above estimate of $T_0$ taken for granted, 
this limit must be rejected physically,
because we hardly accept $\tau_0$ as long as 
$\hbar/T_0 > 10^5$~second on physical grounds.
Hence 
it is valid to use
the single length scale (\ref{lam0}) in free space,
since we can consistently assume
\begin{equation}
T_0\tau_0\ll 4\pi \hbar
\label{T0tau0llhbar} 
\end{equation}
without invalidating the model.

%
%Specifically, the collapse time $\tau_0$ must be 
%longer than characteristic time scales of microscopic phenomena, but
%short enough to ensure that 
%we have never observe such 
%a problematical superposition of 
%the ``classical'' states with ``distinct'' spatial configurations.
%Needless to say,
%we should assess the distinctness of position
%$\bar{\bf r}$ in terms of 
%$|\delta {\bf r}|$.
%By way of illustration, 
%the density profile $\rho(x,t)$ of a free wave packet
%expected in the modified dynamics is 
%schematically displayed in Fig.~\ref{fig:free2}.

\section{\label{sec:miod}Motion in one dimension}\label{sec:4}

\subsection{Model}\label{sec:4-1}
%Collapse: Cell Division}

To give shape to the model,
we investigate motion of a particle in one dimension.
The unmodified Schr\"odinger equation may be written 
in a dimensionless form,
\begin{equation}
{\rm i} \frac{\partial}{\partial \tilde{t}} \Psi
=-\frac{\partial^2}{\partial \tilde{x}^2} \Psi +V(\tilde{x}),
\label{H1dimless}
\end{equation}
where $\tilde{t}= T_0 t/\hbar$ and $\tilde{x}= x/\lambda_0$.
As discussed above, 
for macroscopic classical bodies, 
the potential $V(\tilde{x})$ 
should be slowly varying, $|\nabla V|\ll |V|$.
In practice,
the time scale $T_0/\hbar$
of (\ref{H1dimless})
should be substantially slower than $\tau_0$,
as noted in (\ref{T0tau0llhbar}).
Nevertheless, in this section,
we assume $T_0\tau_0=\hbar$ and $\tilde{t}= t/\tau_0$,
for simplicity,
to illustrate how the probabilistic and deterministic
time evolutions compete with each other.
%
%illustrating how the probabilistic and deterministic
%
%The main objective of this section is to demonstrate 
%by numerical simulation
%that qualitative features resulting from the proposed dynamics
%fit in with the classical-mechanical world view.
%on the basis of the kinematics of quantum mechanics.

%The results are obtained numerically,
%and shown {on a computer display screen}.
%This is a salient feature of our model.
%since not all the other interpretations of QM 
%have attempted and achieved it specifically
%for some practical reason or other. 

%stationary 
%the two time scales, $\hbar/T_0$ and $\tau_0$.
%the former outweigh
%
%******

It is generally difficult
to evaluate the localization functions analytically
for an arbitrarily given state (cf. Appendix~\ref{ap:vari}),
so that we discuss only the simple case that 
the state $\Psi(x)$ is divided by collapse
into the left and right halves at a position $x_0$.
To describe the collapse, 
let us make use of the two trial functions 
for the localization functions 
as follows (cf. Fig.~\ref{fig:Pn}),
\begin{equation}
P_L^{x_0}(x)=
\left\{
\displaystyle
\begin{array}{cc}
1, & x-x_0 \le -2\lambda \\
\cos\frac{\pi}{8\lambda}{(x-x_0+2\lambda)},
& |x-x_0|\le 2\lambda
\\
0,& 2\lambda \le x-x_0
\end{array}
\right.
\label{Plx0}
\end{equation}
and
\begin{equation}
P_R^{x_0}(x) =
\left\{
\begin{array}{cc}
0, & x-x_0\le -2\lambda \\
\cos
\frac{\pi}{8\lambda}
{(x-x_0-2\lambda)},
&|x-x_0|\le 2\lambda \\
1,& 2\lambda\le x-x_0
\end{array}
\right.
\label{Prx0}
\end{equation}
with which we obtain
\begin{equation}
\Delta E=
\frac{1}{2m}
\left(
\frac{\pi \hbar}{8\lambda}
\right)^2
\int_{x_0-2\lambda}^{x_0+2\lambda}
\rho(x) {\rm d} x.
\label{Delta E}
\end{equation}
The entropy change is approximately given by
\begin{equation}
 \Delta S\simeq -
w_R \log w_R-w_L \log w_L,
\label{DeltaSsimeq}
\end{equation}
where
\[
w_L=
\int_{-\infty}^{\infty} (P_L^{x_0} (x))^2 \rho(x) {\rm d} x
\]
and $w_R=1-w_L$ are the probabilities to 
result in $P_L^{x_0}\Psi$ and $P_R^{x_0}\Psi$, respectively.

In order to fix $x_0$ and $\lambda$, 
we have to evaluate $S'$ in (\ref{S'0})
for the given density $\rho(x)=|\Psi(x)|^2$ at every moment.
Nevertheless, to reduce 
the numerical task,
we assume to set $\lambda=\lambda_0/8$ arbitrarily, 
and choose $x_0$ so as to minimize
$\Delta F\equiv \Delta E-T_0 \Delta S$ with (\ref{DeltaSsimeq})
in anticipation that
the qualitative features shown below would not be 
severely modified by the assumption.
In effect, 
we are interested in the interplay
of the stochastic and deterministic evolution.
%how optimally 
%whether the parameters 
%%to characterize the localization functions 
%are optimized,
%for simplicity, 
%we assume to set $\lambda=8\lambda_0$ arbitrarily, 
%and choose $x_0$ so as to minimize
%$\Delta F\equiv \Delta E-T_0 \Delta S$
%with (\ref{DeltaSsimeq}).
%independently of $\rho(x)=|\Psi(x)|^2$, 
%we adopt (\ref{Plx0}) and (\ref{Prx0})
%for fixed $l=\lambda_0$, (\ref{lam0}).
%Then, given the density $\rho(x)=|\Psi(x)|^2$, 
%we choose ${x_0}$
%to minimize
%\begin{equation}
%\Delta F\equiv 
%\Delta E-T_0 \Delta S,
%\label{DeltaF} 
%\end{equation}
%while checking if $\Delta F$ is negative indeed.
%and the Boltzmann constant $k_{\rm B}=1$.
In what follows,
we set $\hbar =1$, while assuming
$T_0=1$ and $\tau_0=1$ despite (\ref{T0tau0llhbar}).

\subsection{Results}\label{sec:4-2}

%It is straightforward to implement
%the above scheme numerically
%for one dimensional systems.
%described by (\ref{H1dim}).
%In effect, 
%the figures shown below are drawn with data output by
%a program, with various inputs as specified in the captions.

\begin{figure*}
\includegraphics[width=0.49\textwidth]{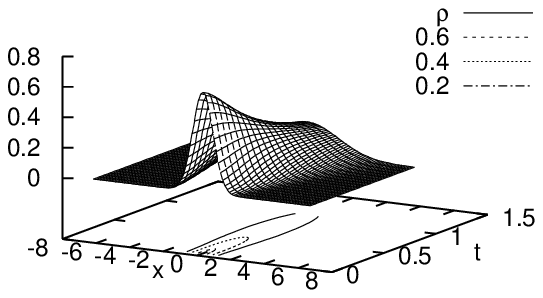}
\includegraphics[width=0.49\textwidth]{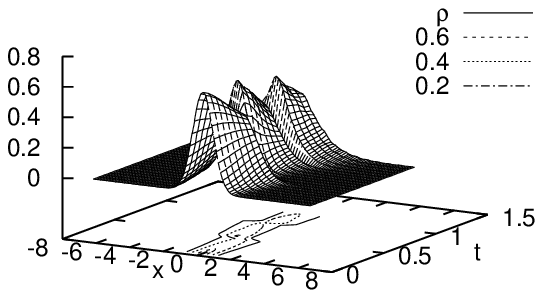}
%\includegraphics[width=0.49\textwidth]{output3}
%\includegraphics[width=0.49\textwidth]{output2}
% Here is how to import EPS art
\caption{
\label{fig:free2}
Time evolution of 
the wave packet density $\rho(x,t)$ of a free particle with
$2m=1$, starting from 
the Gaussian $\Psi(x,0)\propto {\rm e}^{-(x/\delta)^2}$ with
%$\delta=\lambda_0/6$.
$\delta=\lambda_0/3$
(cf. $\lambda_0=3.5$).
%(cf. $\lambda\simeq 0.4$).
Without (left) and with (right) collapse.
Two stochastic collapses are recognized.
In our model,
the intermittent collapses keep the wave packet from diffusing,
while setting it in Brownian motion.
%Note that collapse is not put in by hand, but triggered autonomously.
}
\end{figure*}
\begin{figure*}
\includegraphics[width=0.49\textwidth]{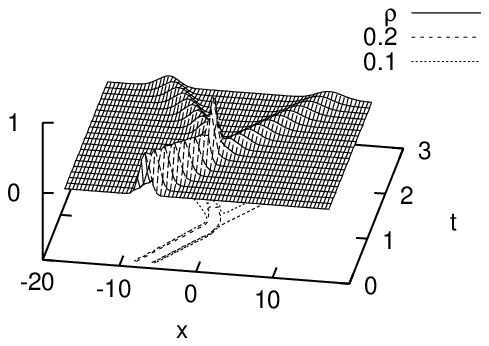}
\includegraphics[width=0.49\textwidth]{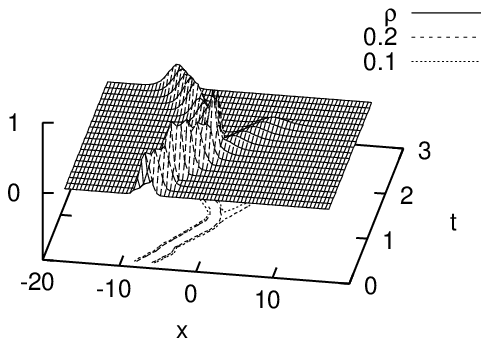}
%\includegraphics[width=0.49\textwidth]{output7}
%\includegraphics[width=0.49\textwidth]{output6}
% Here is how to import EPS art
\caption{\label{fig:tunnel}
Quantum tunneling through a potential barrier at $x\simeq 0$.
The density $\rho(x,t)$ of a particle with $m=1$ in motion, evolving from
$\Psi(x,0)\propto {\rm e}^{-(x/\delta)^2+{\rm i}p_0 x_0}$
with $\delta=\lambda_0/2$,
$x_0=-8$ and $p_0=8$.
The potential is
\(
 V(x)\simeq 0.08 {\rm e}^{-4 x^2}
\).
Without (left) and with (right) collapse.
Two collapses before and after reflection(transmition) are recognized.
%The latter erases the transmitted component.
%the abrupt disappearance of 
%due to the second collapse,
%exemplifies ``spooky action at a distance''. 
Tunneling failed, and the wave packet is reflected.
Aside from a digression over the barrier,
the wave packet traces an approximately well-defined trajectory.
}
\end{figure*}
To begin with, 
we give results for a free particle, $V(x)=0$,
in Fig.~\ref{fig:free2}.
Here and below,
the results for the density $\rho(x,t)$
with and without collapse are compared with each other
to make the non-trivial collapse effect clear.
As shown by contours on the base ($x$-$t$ plane)
of Fig.~\ref{fig:free2},
the collapsing wave packet traces a zig-zag world line.
To a greater or lesser extent,
the discontinuity is a general and non-trivial feature
to be revealed in our results.
This is made conspicuous when 
we should otherwise, without collapse,
get involved in a controversial superposition state.
As a typical case, 
results for a moving particle impinging on
a potential barrier at $x\sim 0$ are given in
Fig.~\ref{fig:tunnel}.

%which are generally expected in the presence of a potential. 
%we calculate motion of a wave packet under a potential. 

\begin{figure*}
\includegraphics[width=0.49\textwidth]{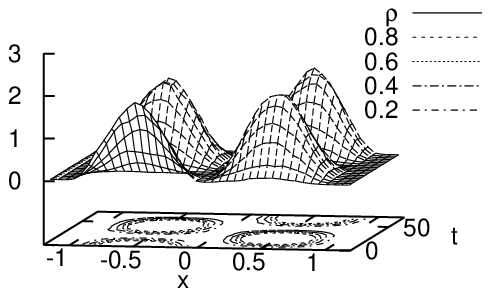}
\includegraphics[width=0.49\textwidth]{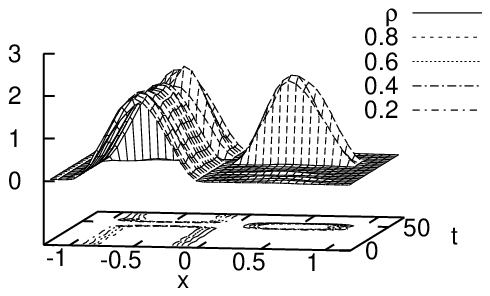}
\caption{
\label{fig:doublewell}
Quantum oscillation 
in a double well potential ($m=2$). 
Without (left) and with (right) collapse.
There are potential walls at $x=\pm 1$, and 
a square potential of the width and the height of 0.1 at $x=0$.
%$l=\lambda_0/2$.
The sinusoidal continuous oscillation apparent in the left figure
is suppressed by collapse.
On the time scale of the figure,
individual collapse are not conspicuous.
%that the time scale of the oscillation is elongated by collapse.
In the right figure,
we observe that
the particle is found to be either in $x>0$ or $x<0$,
instead of being in their superposition.
The particle position changes discontinuously from time to time
 at random.
In a conventional phrase,
the particle behaves
as if it is ``measured'' with the frequency $\gamma_0$.
%the world line (contour lines) 
%of the continually collapsing state 
%switches on and off at random.
%like a telegraph.
% with an elongated time scale.
%The particle is found in either of 
%Thus we observe
%a ``chiral'' state resulting from
%an arbitrary superposition of
%parity eigenstates.
}
\end{figure*}
\begin{figure*}
\includegraphics[width=0.49\textwidth]{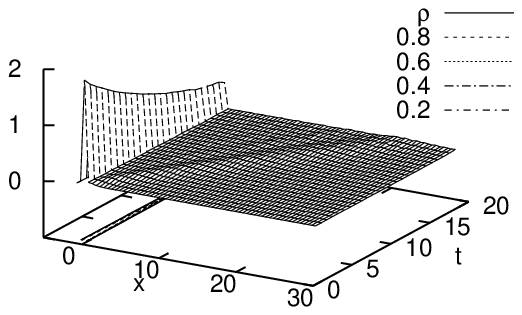}
\includegraphics[width=0.49\textwidth]{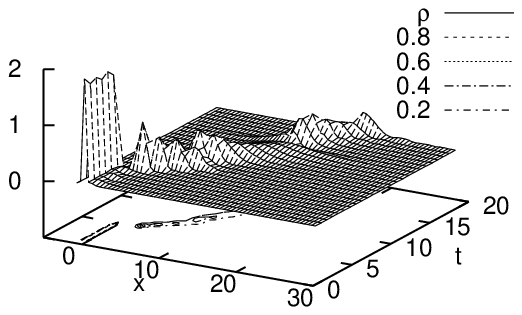}
\caption{
\label{fig:decay}
Quantum decay of a particle $m=2$ trapped in a potential well.
The potential well $-1\le x<0$ is that of Fig.~\ref{fig:doublewell} 
with the wall at $x=1$ removed.
Without (left) and with (right) collapse.
The wave packet is literally trapped until it goes off
at a definite, but unpredictable, instant.
%Wave function collapse causes the trapped
%wave packet to escape 
%at a definite (but unpredictable) instant.
%The decay obeys an exponential law.
%The decay rate does not depend on $\tau_0$, in practice.
}
\end{figure*}
Results for the system in a double well potential
are shown in Fig.~\ref{fig:doublewell},
for which we varied $x_0$ for fixed 
$\lambda=\lambda_0/16$,
%$l=\lambda_0/2$,
as the constraint allows 
the width $\lambda$ to be smaller in this case.
We find that
even $\lambda=\lambda_0/32$ does not change the results
qualitatively, for
the wave function tends to collapse around $x\simeq 0$.
The suppression of a coherent quantum motion by frequent ``measurements''
is generally known as the quantum Zeno effect\cite{ms77,pe80,kr80}.
This is demonstrated in Fig.~\ref{fig:doublewell}.
We are also interested in a related problem of quantum decay from a
potential well.
To realize such, 
we solved a simple model as indicated in Fig.~\ref{fig:decay}.
It is noted that the unstable state trapped 
in the potential barrier
looks stationary
until it decays spontaneously.

\begin{figure*}
\includegraphics[width=0.49\textwidth]{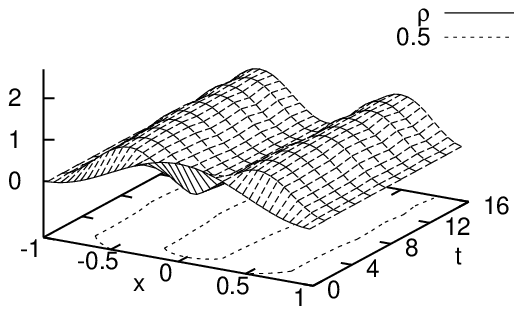}
\includegraphics[width=0.49\textwidth]{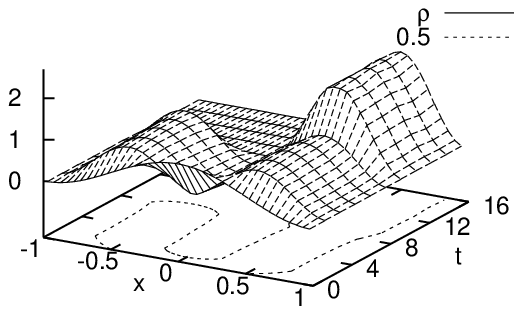}
\caption{
\label{fig:doublewell2}
Quantum ``measurement'' caused by a time-dependent potential.
A partition wall is adiabatically 
inserted at the center of 
a bound state ($2m=1$) in
the potential well $-1\le x\le 1$.
Without (left) and with (right) collapse taken account of.
The potential is the same as in Fig.~\ref{fig:doublewell},
except that the height of the central square potential 
is increased as $v_{\rm height}=0.1t$.
The wave packet is finally found 
in either of the left or right partition.
}
\end{figure*}
Lastly, to mimic situations encountered in position measurement,
wave packet reduction induced by a time-dependent potential is
displayed in Fig.~\ref{fig:doublewell2},
which shows that gradual increase of the external potential
may cause a sudden change of state.
To obtain Fig.~\ref{fig:doublewell2},
we varied $\lambda$ for $x_0=0$.
This behavior corresponds to the transition
from (\ref{phisigtophisig}) to (\ref{Psitophisigma}).
In general,
we associate the ``measurement'' of a quantum particle
with such an unordinary leap 
caused by a specially prepared external potential.

These results demonstrate that
the effect of collapse becomes conspicuous when 
the linear width $\delta$ of the wave packet exceeds 
the characteristic length scale $\lambda_0$,
and that the collapse hardly affects the wave packet 
when $\delta\ll \lambda_0$.
Therefore, as discussed in the previous sections,
we obtain a picture in that 
quantum (wave) mechanics obtains in the one limit 
$\lambda_0\rightarrow\infty$,
while classical (particle) mechanics (with a definite path)
emerges in the other limit
$\lambda_0\rightarrow 0$.

\section{Discussion}\label{sec:discussions}

According to the model,
wave function reduction causes
the density matrix of the pure state $\Psi({\bf r})$ to evolve as 
\begin{equation}
\rho({\bf r},{\bf r'})=
 \Psi({\bf r})\Psi^*({\bf r'})
\longrightarrow
{\mit\Delta ({\bf r}, {\bf r'})}
%\left(\sum_n P_n({\bf r})P_n^*({\bf r'})\right)
 \Psi({\bf r})\Psi^*({\bf r'}),
%\equiv T[ \Psi({\bf r})\Psi^*({\bf r'})
\label{rhorr'} 
\end{equation}
where
\begin{equation}
\mit\Delta ({\bf r}, {\bf r'})
\equiv 
\sum_n 
P_n({\bf r})P_n^*({\bf r'})
\label{mitdeltarr'}
\end{equation}
equals 1 for ${\bf r}={\bf r'}$ because of (\ref{sumnPnP2=1}), and
vanishes for $|{\bf r}-{\bf r'}|\gg \lambda$.
Hence, by postulate,
the diagonal elements
$\rho({\bf r},{\bf r})=\rho({\bf r})$
are invariant upon collapse.
As the functions $P_n$ are determined by $\rho({\bf r})$, 
so is $\mit\Delta ({\bf r}, {\bf r'})$.
Therefore, 
(\ref{rhorr'}) generally represents a non-linear transformation.
Nevertheless, in many practical cases,
the localization functions may be determined 
by the external potential $V({\bf r})$,
as in Figs.~\ref{fig:doublewell} and \ref{fig:doublewell2}.
In fact, this would be the case 
when we can neglect wave function collapse except
in measurement-related situations.
In such cases, 
we obtain a linear equation,
referring to Fig.~\ref{fig:chart}, 
%To put (\ref{rhorr'})  more precisely,
%we obtain (cf. Fig.~\ref{fig:chart})
\[
 \rho_{t+{\rm d}t}
%({\bf r},{\bf r'})
=\left(1-\gamma_0 {\rm d} t\right)
\left(
\rho_t+
 \frac{{\rm i}}{\hbar}[
\rho_t
%({\bf r},{\bf r'})
,H]_-{\rm d} t
\right)
+\gamma_0{\rm d} t
\mit\Delta ({\bf r}, {\bf r'})
\rho_t,
%({\bf r},{\bf r'})
\]
or
\begin{equation}
 \frac{{\rm d} \rho}{{\rm d} t}= 
 \frac{{\rm i}}{\hbar}[\rho,H]_-
-\gamma_0\left(1-\mit\Delta
\right)\rho.
\label{delrhodelt}
\end{equation}
%This obtains also in
%the other case of interest.
%In the limit $|x-x'|\ll l$, 
%$x\simeq x_0$,
%(\ref{Plx0}) and (\ref{Prx0}) give
%\begin{eqnarray}
%1- {\mit\Delta (x, x')}
%&\simeq &
%\frac{\pi^2}{2l^2}
%\left(x-x'\right)^2,
%\label{1-Delxx'}
%\end{eqnarray}
%with 
%\(
% l=\sqrt{\frac{\pi}{4\log 2}}\lambda_0
%\)
%for $w_L=w_R=1/2$.
The equation of motion (\ref{delrhodelt})
has
the same form as that
investigated by Joos and Zeh
to assess the effect of 
decoherence through interaction with the environment\cite{jz85}.
%Not to mention numerical details, 
%the contribution (\ref{1-Delxx'}) due to collapse would be completely 
%outweighed in practice by the extrinsic decoherence.

%\[
% \cos(x'-x)=
% \cos(-x)-x' \sin(-x)-\frac{x'^2}{2} \cos(-x)
%\]
%\[
% \sin(x'-x)=
% \sin(-x)+x' \cos(-x)-\frac{x'^2}{2} \sin(-x)
%\]
%\[
% \cos(x'-x)=
% \cos(x')\cos(x)-
% \sin(x')\sin(x)=
% \cos(-x)-x' \sin(-x)-\frac{x'^2}{2} \cos(-x)
%\]
%\[
% \sin(x'-x)=
%\sin(x')\cos(x)-\cos(x')\sin(x)=
% \sin(-x)+x' \cos(-x)-\frac{x'^2}{2} \sin(-x)
%\]
%
%\begin{eqnarray}
% \cos(x)\cos(x+x')
%+ \sin(x)\sin(x+x')
%&=&
% \cos x(\cos x\cos x'-\sin x\sin x')
%+\sin x(\sin x\cos x'+\cos x \sin x')
%\nonumber\\
%&=&\cos x'
%\end{eqnarray}

%Generally speaking,
At the statistical level of description,
the collapse (\ref{rhorr'}) does not affect
the density matrix $\rho({\bf r},{\bf r}')$ unless 
it has the longer ranged correlation than 
the correlation length $\lambda$ of (\ref{mitdeltarr'}).
%spatial correlation of the correlation length 
%extending over a
%length 
For example, 
for free particles in thermal equilibrium
at temperature $T$, we have
\begin{equation}
 \rho({\bf r},{\bf r'})\simeq \lambda_T^{-3/2}
\exp\left(-\frac{\pi|{\bf r}-{\bf r'}|^2}{\lambda_T^2}
\right), 
\label{rhorr'T}
\end{equation}
where $\lambda_T$ is the thermal de Broglie wavelength at $T$.
In order for (\ref{rhorr'}) to affect
the statistical results based on (\ref{rhorr'T}),
we should assume $\lambda_T\gg \lambda_0$, or the physical
temperature must be 
extremely low, $T\ll T_0$.
Consequently, in the statistical ensemble,
the phenomenological effects of collapse 
apparent at the individual level can be completely masked 
by environmental decoherence\cite{gf90,te93,gjkksz96,zu03,tvg95}.
%since in almost all practically relevant situations
%the opposite limit $T\gg T_0$ surely holds.
In other words,
the discontinuity due to collapse as indicated
in Figs.~\ref{fig:free2} and \ref{fig:doublewell2}
would be averaged out at the statistical level.
%in the statistical description based on $\rho({\bf r},{\bf r}')$.
%This is why 
%we should put more emphasis on the wave function
%than the density matrix.

%The former determines the latter uniquely, but not vice versa.
%The latter is uniquely obtained by the former

Last but not least, 
the theory discussed throughout the paper is essentially 
non-relativistic.
This is evident from a special role we assigned to time $t$.
It is still noteworthy that the model 
formulated with $\rho({\bf r})$
is Galilei invariant.
The conflict of the instantaneous collapse
with special relativity has been elucidated\cite{aa81}.
%While the statistical evolution equation (\ref{delrhodelt}) 
%cannot be used to send superluminal signals,
%the non-linearity of (\ref{rhorr'}) 
%may pose a problem in connection with relativity\cite{gi89}.
%At present,  we have no idea
%to make our model Lorentz invariant.
In this regard, we only note 
a consistent if naive attitude 
to fit with the present model,
that is, to assume a special frame of reference,
as noted by Eberhard\cite{eb78} and Bell\cite{bell87bqft},
in which the collapse dynamics applies specifically\cite{noteadded}.

\section{Conclusion}\label{sec:conclusion}

In this paper,
we addressed ourselves to the problem of
how the classical world as we experience it emerges from 
the underlying laws of quantum mechanics.
We investigated the motion of a quantum particle
on the basis of the proposed modified dynamics of wave function
collapse.
In effect, the proposal 
may look complicated in practice,
owing to the task of solving for the localization functions,
but the model is simple in principle,
as summarized in Fig.~\ref{fig:chart}.
%The model dynamics is 
We describe the collapse dynamics 
in terms of physical quantities such as 
energy, entropy, and density in real space.
The model contains two material-independent constants 
$\tau_0=\gamma_0^{-1}$ and $T_0$.
It is suggested on the analogy of
the second law of thermodynamics
in order to destabilize
those linear superpositions of 
spatially separated wave packets,
which are hardly interpretable
from an ordinary, classical, realistic standpoint.
On the basis of the proposal,
the numerical results are obtained as displayed in
Figs.~\ref{fig:free2}-\ref{fig:doublewell2}.
%These results are sought for
%in order to make it 
%more complete the mechanistic view of the world,
%without falling back on such technical 
%notions common in interpreting QM, 
%as measurement, observation, information, etc.
%
The model reproduces classical mechanics as a limit, 
yet it does not use the classical limit for its own formulation.
%To compromise quantum/classical dichotomy, 
According to the model,
we obtain a unified mechanical picture in that
the classical picture emerges
on a long and large scale,
$t \gg\tau_0$ and $x\gg \lambda$,
while
the quantum picture applies
in the opposite limits,
$t \ll \tau_0$ and $x\ll \lambda$,
where
$\lambda$
% for a free particle
%defined in (\ref{lam0}) depends on
%the particle mass $m$,
%and is interpreted as 
denotes the natural linear dimension of the wave packet
determined according to the model.
In fact, the length scale $\lambda$ of collapse
%of the localization functions 
depends on circumstances,
and it can be made arbitrarily small.
This is particularly the case for a microscopic particle
subjected to measurement-related situations,
where one can devise an apparatus to 
make $\lambda$ as small as one likes, in principle.
%The modified dynamics is implemented such that
%the non-trivial classical-quantum crossover
%effect due to wave function collapse
%would not be discernibly reflected
%in a statistical ensemble of the collapse outcomes
%except through the quantum interference phenomena
%between the collapse outcome states.

%in a statistically averaged results.
%because it would be completely 
%hidden in the presence of 
%overwhelmingly relevant 
%environmental disturbances
%under normal conditions.

\appendix
%\section{Equations for Localization Functions}
%\section{Analytical Discussions; Mathematical Expressions for Localization Functions}
%\section{Equations for Localization Functions}

\section{Variational Principle 
}\label{ap:vari}

In principle, it is possible to write down 
the equations for $P_n({\bf r})$, that is, 
the equations to determine the collapse outcome states $\Phi_n({\bf r})$.
However, it is generally difficult to solve
them for an arbitrary state $\Psi({\bf r})$ analytically.
In this appendix, 
we investigate the simplest tractable case where 
a wave packet is justly halved with equal probability by collapse.
%we investigate a simple tractable case.
% where 

We obtain the localization functions $P_1$ and $P_2$
under the condition that
the collapse probabilities to the two outcomes are equal,
viz., $w_1=w_2=1/2$.
To this end,
the auxiliary function $\theta({\bf r})$,
ranging from 0 to $\pi/2$,
is introduced by 
$P_1=\cos \theta$ and $P_2=\sin \theta$
to take account of the constraint $P_1^2+P_2^2=1$.
%The function $\theta({\bf r})$ varies from 0 to $\pi/2$ continuously.
%and has to be satisfy
%\[
% w_1=w_2=\int {\rm d}{\bf r} \rho({\bf r}) \cos 2\theta =0
%\]
%
In general, the collapse outcomes $P_n\Psi$ are not orthogonal,
or the overlap integral
\begin{equation}
 p=\langle \Psi|P_1P_2|\Psi \rangle
=\frac{1}{2}\int {\rm d}{\bf r} \rho({\bf r}) \sin 2\theta({\bf r})
\label{overlap-p} 
\end{equation}
does not vanish.
Hence
the entropy production by the collapse is given by
\[
\Delta S= -
\left(\frac{1}{2}+p\right)\log\left(\frac{1}{2}+p\right)
-
\left(\frac{1}{2}-p\right)\log\left(\frac{1}{2}-p\right),
\]
which takes the maximum value $\log 2$ for 
$p=0$,
while the minimum is $\Delta S=0$ for $p=1/2$.

%
%
%Hence the density matrix
%%is 
%%\[
%% \rho({\bf r},{\bf r'})=
%%\Psi({\bf r})^*
%%\left( P_1({\bf r}) P_1({\bf r'})
%%+P_2({\bf r}) P_2({\bf r'})
%%\right)\Psi({\bf r}'),
%%\]
%%which 
%has to be diagonalized to estimate the entropy. 
%In this case,
%we obtain
%\[
% \rho({\bf r},{\bf r'})=
%\sum_{\sigma=\pm}\lambda_\sigma \phi_\sigma({\bf r})\phi_\sigma({\bf r}'),
%\]
%where
%\begin{eqnarray}
% \lambda_\pm &=&\frac{{1\pm p}}{2},
%\nonumber\\
% \phi_\pm &=&\frac{P_1\Psi \pm P_2\Psi}{\sqrt{2(1\pm p)}},
%\nonumber
%\end{eqnarray}
%and the overlap integral
%\begin{equation}
% p=\langle \Psi|P_1P_2|\Psi \rangle
%=\frac{1}{2}\int {\rm d}{\bf r} \rho({\bf r}) \sin 2\theta({\bf r}).
%\label{overlap-p} 
%\end{equation}
%Consequently, the entropy due to the collapse is given by
%\begin{equation}
%S= -\sum_{\sigma=\pm} \lambda_\sigma \log \lambda_\sigma,
%\label{entropySlambda} 
%\end{equation}
%which takes the maximum value $\log 2$ for $p=\langle \Phi_1|\Phi_2\rangle=0$.
%

In order to achieve the localization of wave function, 
to maximize $\overline{\rho^2}$ in (\ref{rho2})
is more easily implemented than to 
work with $(\delta {\bf r})^2$, (\ref{delr^2}).
Hence, to start with, let us maximize
\begin{equation}
\overline{\rho^2}
=
 2 \int (\sin^4\theta+\cos^4\theta)
\rho({\bf r})^2 {\rm d}{\bf r}
\label{rho2=2int}
\end{equation}
%\begin{eqnarray}
%\overline{\rho^2}
%&=&
% 2 \int (P_1^4+P_2^4)
%\rho({\bf r})^2 {\rm d}{\bf r}
%\nonumber\\
%&=&
% 2 \int (\sin^4\theta+\cos^4\theta)
%\rho({\bf r})^2 {\rm d}{\bf r}
%\nonumber
%\end{eqnarray}
under (\ref{T0DeltaS=DeltaE}), 
for which
\begin{equation}
\Delta E=
\frac{\hbar^2}{2m} 
\int 
\left(
\nabla \theta
\right)^2
\rho({\bf r}) {\rm d}{\bf r}.
\label{varepsilon0} 
\end{equation}
%
%which reads
%\begin{equation}
%\Delta E
% \varepsilon  =T_0 S
%\label{varepsilon=}  +++++++++++++++++++
%\end{equation}
%%\begin{equation}
%% \varepsilon  \simeq T_0 \log 2, ***
%%\end{equation}
%where 
%%\begin{equation}
%% \varepsilon=
%%\frac{\hbar^2}{2m} 
%%\int 
%%\left(
%%\nabla \theta
%%%\frac{{\rm d}\theta}{{\rm d}x}
%%\right)^2
%%\rho({\bf r}) {\rm d}{\bf r}.
%%\label{varepsilon} 
%\end{equation}
In terms of a Lagrange multiplier $\lambda'$,
the equation for $\theta$ is obtained by the variation principle,
%\[
% \frac{\delta}{\delta \theta} (\overline{\rho^2} 
%-\lambda'  \varepsilon
%)=0,
%\]
\[
 \frac{\delta}{\delta \theta} (\overline{\rho^2} 
+
\lambda' (T_0 \Delta S-  \Delta E)
)=0,
\]
from which 
we obtain the equation
\begin{equation}
\lambda'
\left(
\frac{\hbar^2}{m} 
%\left(
%\rho\nabla^2 \theta+\nabla \theta\nabla \rho
%++
\nabla( \rho \nabla \theta)
-T_0 \frac{\rho}{2}\log \frac{1+p}{1-p}\cos 2\theta
\right)
=
%\right)=
2 \rho^2 \sin 4\theta.
\label{lam'nab2the+nabtenabrho}
\end{equation}
In particular, we are interested in the case $p\simeq 0$
and $|\nabla\rho/\rho|\ll |\nabla \theta|$,
where the model predicts the non-trivial result
that the collapse outcomes $P_1\Psi$ and $P_2\Psi$ are 
distinctly different from each other,
$\langle \Phi_1|\Phi_2\rangle \simeq 0$.
%
%
%Therefore, neglecting the second term in the right-hand side of 
%(\ref{lam'nab2the+nabtenabrho}),
%
%
%... $p\simeq 0$,  ...
%$\hbar^2|\nabla^2 \theta|/m\gg T_0$
%
%To discuss
%the case $|\nabla\rho/\rho|\ll |\nabla \theta|$
%of our interest concretely, let us 
%$\rho$ can be regarded as constant over 
%the length scale where $\theta$ varies appreciably, and 
To discuss this case further in depth, 
regarding $\theta$ as a function of $x$ for simplicity,
%we can neglect the second term in
%the left-hand side of (\ref{lam'nab2the+nabtenabrho})
%compared with the first term.
%Moreover, 
%taking the $x$ axis along the direction of $\nabla \theta$,
we obtain the one-dimensional equation
\begin{equation}
4\lambda^2 \frac{{\rm d}^2\theta}{{\rm d}x^2}=\sin 4\theta, 
\label{pendulum}
\end{equation}
where
\(
 \lambda=\sqrt{\hbar^2 \lambda'/8m \rho}.
\)

\begin{figure}
\begin{center}
\includegraphics[width=0.4\textwidth]{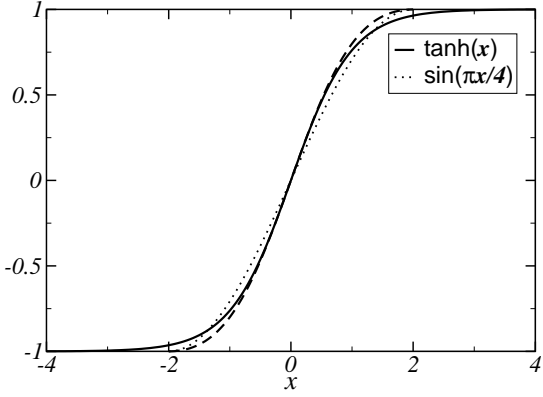}
\caption{
The localization function $2 P_2(x)^2-1=- \cos 2\theta$ 
is shown as a function of $x$.
The solid and dashed curve are the solutions of 
(\ref{pendulum}) and (\ref{pendulum2}) for $\lambda=1$, respectively. 
The dotted curve is a sine curve, 
which is added for reference.
}
\label{fig:Pn}
\end{center}
\end{figure}

The equation of a pendulum (\ref{pendulum}) can be solved analytically.
The solution with a proper boundary condition is given by 
\begin{equation}
\cos 2\theta =  -{\rm sn}(x/\lambda,1)
=-\tanh (x/\lambda),
\label{solutionsn}
\end{equation}
where ${\rm sn}(u,k)$ is
the Jacobian elliptic function.
%for and $k=1$. We implicitly chose 
The origin of the $x$ axis
has to be consistently chosen to give $w_1=w_2=1/2$, i.e., by
\begin{equation}
\int 
%{\rm sn}(x/\lambda)
\tanh(x/\lambda)
\rho {\rm d}{\bf r}=0.
\label{inttanh=0} 
\end{equation}
The solution (\ref{solutionsn}) 
for $\lambda=1$ 
is shown in Fig.~\ref{fig:Pn}
with the solid curve.
Note that the curve essentially represents 
the localization function $P_2(x)^2$, because 
\[
 P_{1,2}(x)^2=\frac{1\pm \cos 2\theta}{2}.
\]
To estimate the energy cost 
(\ref{varepsilon0}) to fix $\lambda$, 
we can make use of `energy conservation' of
the pendulum motion (\ref{pendulum}),
\begin{equation}
8\lambda^2 \left(\frac{{\rm d}{\theta}}{{\rm d}x}
\right)^2+V(\theta)
%\frac{1}{4} \cos 4\theta = 
=V(0),
%\frac{1}{4}
\label{energy-conservation} 
\end{equation}
where 
\begin{equation}
V(\theta)=\frac{1}{4}\cos 4\theta.
\label{Vtheta}
\end{equation}
From the last equations, we obtain
\[
4 \lambda^2\left(\frac{{\rm d}{\theta}}{{\rm d}x}
\right)^2
=P_1(x)^2 P_2(x)^2,
\]
hence, 
\begin{equation}
\Delta E=
\frac{\hbar^2 q}{8 m \lambda^2},
\label{varep=q}
\end{equation}
where
\begin{equation}
 q=
\int P_1^2(x) P_2^2(x)
\rho {\rm d}{\bf r}
\simeq 
\int 
{\rm e}^{-x^2/\lambda^2}
\rho {\rm d}{\bf r}.
\end{equation}
With respect to the last approximation, we should note
$\int P_1(x)^2P_2(x)^2 {\rm d}x=2\lambda$,
while $\int {\rm e}^{-x^2/\lambda^2}{\rm d}x=\sqrt{\pi}\lambda$,
and
$P_1(x)^2P_2(x)^2\simeq {\rm e}^{-x^2/\lambda^2} \simeq 1-(x/\lambda)^2$
for $|x|\ll \lambda$.
Similarly, we may write (\ref{overlap-p}) as
\begin{equation}
 p 
%\int P_1(x) P_2(x)
%\rho {\rm d}{\bf r}
\simeq 
\int {\rm e}^{-x^2/2\lambda^2}
\rho {\rm d}{\bf r},
\end{equation}
with which $\rho({\bf r})=|\Psi({\bf r})|^2$ 
must meet $p\ll 1$ as well as (\ref{inttanh=0}) for consistency.
Given such $\rho({\bf r})$,
the parameter $\lambda$ for $P_n({\bf r})$ is determined 
by $\Delta E \simeq T_0\log 2$, as in (\ref{lambdasimeqlambda0q}).
In consequence, we obtain
the collapse outcomes $\Phi_n=\sqrt{2}P_n \Psi$.
Note in particular that
$P_n$ becomes the step function
in the limit $\lambda\rightarrow 0$,
e.g., 
as a result of the low density at $x=0$,  $\rho(x=0)\simeq 0$.
Then
 we recover the Born's-rule probabilities
$w_n=|\langle \Phi_n|\Psi\rangle|^2$
together with the non-overlapping collapse outcomes $\Phi_n$,
owing to $P_n^2\simeq P_n$.
As is clear from the qualitative discussion in 
Section~\ref{sec:3-2},
this feature is generic and 
not specific to the assumption $w_1=w_2$.

%The assumption $p\simeq 0$ requires 
%that the width $\lambda$ is smaller than 
%$\rho({\bf r})$.

%\section{entropy}

%Lastly, it is noteworthy that 
%we can equally make use of 
%the other expressions than $\overline{\rho^2}$ to
%realize localization.

In a similar manner, 
we can fix $P_n$ by maximizing (\ref{S'0})
instead of $\overline{\rho^2}$, (\ref{rho2=2int}).
%\begin{equation}
% S'= \sum_n w_n \int |\Phi_n|^2
%(\log |\Phi_n|^2 -1 ) {\rm d}{\bf r}.
%\label{S'}
%\end{equation}
To obtain $\Phi_n$ from $\Psi$,
%the spontaneous collapse $\Psi\rightarrow \Phi_n$ 
our task is to maximize
%our collapse rule is simply stated
%that 
%to 
%the spontaneous collapse $\Psi\rightarrow \Phi_n$ 
%occurs to maximize 
%$S'+\beta(T_0S-E)$. Then our task is to maximize
\[
S'+\beta( T_0 \Delta S-\Delta E),
\]
%\[
%E-T_0S -T'S'
%\]
where $E={\rm Tr} (\hat{\rho} \hat{H})$,
$S=-{\rm Tr} (\hat{\rho} \log \hat{\rho})$, 
and $\beta$ is the Lagrange multiplier to 
keep the postulated constraint $\Delta (E-T_0S)=0$.
In the same approximations as above, 
in place of 
%(\ref{lam'nab2the+nabtenabrho}) and 
(\ref{pendulum}), we obtain
%\begin{equation}
%4\lambda^2 
%%\beta\frac{\hbar^2}{m} 
%\nabla( \rho \nabla \theta)=
%\rho 
%\sin 2\theta \log \cot \theta,
%\end{equation}
%and
%AAAAAAAAAAAAAA
%In the same approximations as above, in place of (\ref{pendulum2}),
%we obtain
\begin{equation}
4\lambda^2 
\frac{{\rm d}^2\theta}{{\rm d}x^2}=\sin 2\theta \log \cot \theta,
\label{pendulum2}
\end{equation}
where \(4\lambda^2 ={\hbar^2}\beta/{m}\).
A numerical solution $\cos 2\theta$ for $\lambda=1$ 
is shown in Fig.~\ref{fig:Pn}
with the dashed curve.
We find that 
the optimal shape of the localization function $P_n(x)$ is not severely
modified.
% by whichever quantity we use to represent the degree of compact localization.
%we choose to achieve compact localization.
For reference, a sinusoidal function $\sin\pi x/4$ 
for $|x|\le 2$ is also 
shown in Fig.~\ref{fig:Pn}
with the dotted curve.

It is noted that the solution of (\ref{pendulum2}) gives 
a finite value for $|x(\theta=\pi/2)-x(\theta=0)|$,
which is the overlap width of $P_1(x)P_2(x)$.
This is because the `potential'
\(
 V(\theta)=
(\log \sin (2\theta)+\cos(2\theta) \log\cot\theta)/2
\)
for (\ref{pendulum2}) makes the integral
\[
 \int_0^{\pi/2} \frac{{\rm d}\theta }{\sqrt{V(0)-V(\theta)}}
\]
convergent,
%gives a finite value to 
while (\ref{Vtheta})
%$V(\theta)=\cos(4\theta)/4$ for (\ref{pendulum}) 
makes this expression divergent.
Thus we obtain compact support of $\Phi_n$,
which is a favorable feature of the model.
Moreover,
from (\ref{energy-conservation}), we obtain
$|{\rm d}\theta/{\rm d}x| \rightarrow 0$ as $\theta\rightarrow 0, \pi/2$
for the solution of (\ref{pendulum2}),
by which follows the continuity of $\Phi_n(x)$ as well as their
derivatives as a function of $x$.
%To give compact support of $\Phi_n$
%is a favorable feature of the model.
Furthermore, as discussed below,
the expression (\ref{S'0}) 
is generalized adequately
to describe independent collapse of many particle systems.

\section{Generalization to Many-particle systems}\label{ap:many}

In the main text, 
we focused on single-particle problems
of the modified dynamics.
To describe collapse of a many particle system,
we have to generalize the localization functions
to $P_n({\bf r}_1,{\bf r}_2,\cdots, {\bf r}_N)$,
which are to be multiplied with
the $N$ particle wave function
$\Psi({\bf r}_1,{\bf r}_2,\cdots, {\bf r}_N)$.
Accordingly, the integral $\int {\rm d}{\bf r}$
in (\ref{wndef})
is generalized to
$\int {\rm d}{\bf r}_1 {\rm d}{\bf r}_2\cdots  {\rm d}{\bf r}_N$
over $3N$ dimensional space for the $N$ particles.

For $N$ distinguishable particles, one should maximize
\begin{equation}
 S'= \sum_n w_n \int |\Phi_n|^2
\log |\Phi_n|^2 
{\rm d}{\bf r}_1 {\rm d}{\bf r}_2\cdots  {\rm d}{\bf r}_N,
%--  S'= \sum_n w_n \int |\Phi_n|^2(\log |\Phi_n|^2 -1 ) {\rm d}{\bf r},
\label{S'N}
\end{equation}
instead of (\ref{S'0}).
In particular,
when the $N$ particles are independent from each other,
the wave function of the whole system is represented
by the product of wave functions,
\[
\Psi({\bf r}_1,{\bf r}_2,\cdots, {\bf r}_N)
=\prod_{i=1}^N \psi_i({\bf r}_i).
\]
for which one can regard formally that each of the states
$\psi_i({\bf r}_i)$ collapses independently,
\[
\psi_i\rightarrow 
\phi_{n_i}= \frac{1}{\sqrt{w_{n_i}}}P_{n_i}\Psi_i, 
\]
since (\ref{S'N})
is decomposed into a sum of $N$ independent terms,
\begin{eqnarray}
 S'&=&\sum_{i=1}^N 
%S'_i,\nonumber\\S'_i&=&
\sum_{n_i} w_{n_i} \int |\phi_{n_i}|^2 \log  |\phi_{n_i}|^2
{\rm d}{\bf r}_i,
\nonumber
\end{eqnarray}
owing to $w_n= \prod_i w_{n_i}$. 
This is a feature to be expected physically.

For indistinguishable particles,
the symmetry of the wave function $\Psi$
must be respected,
so that
the localization functions
$P_n({\bf r}_1,{\bf r}_2,\cdots, {\bf r}_N)$
must be symmetric under the exchange of the arguments.
For definiteness,
they are approximately given
in terms of $P_n({\bf r})$ for a single particle state
as follows.
The simplest assumption is to regard them simply
as functions of 
the center-of-mass coordinate 
${\bf R}=\sum_i {\bf r}_i/N$,
\[
P_n({\bf r}_1,{\bf r}_2,\cdots, {\bf r}_N)=P_n({\bf R}).
\]
Otherwise,
%the many-particle localization functions 
%$P_n({\bf r}_1,{\bf r}_2,\cdots, {\bf r}_N)$
they may be represented as symmetrized products 
of $P_n({\bf r})$, 
\begin{equation}
 P_n({\bf r}_1,
{\bf r}_2,
\cdots, {\bf r}_N)^2 
=\sum P_{n_1}({\bf r}_1)^2 
P_{n_2}({\bf r}_2)^2
\cdots
P_{n_N}({\bf r}_N)^2,
\nonumber
\end{equation}
where 
$n$ represents a set of integers $(n_1, n_2, ...., n_N)$
%$n=(n_1,n_2,\cdots, n_N)$
and 
the sum in the right-hand side 
is taken over all different permutations of them,
% different suffices $n_1$, $n_2$, $\cdots$, $n_N$.
so that
\begin{eqnarray}
 \sum_n P_n^2=
\prod_{i=1}^N \left(\sum_{n_i}P_{n_i}({\bf r}_i)^2\right)
=1.
\nonumber
\end{eqnarray}
%To fix the localization functions 
For identical particles,
%instead of (\ref{S'N}) which depends explicitly on $N$,
%one may disregard the particle correlation in (\ref{S'N}),
the following expression may equally serve our purpose,
\begin{equation}
 S'= \sum_n w_n \int\rho_n({\bf r})
\log \rho_n({\bf r}){\rm d}{\bf r},
\end{equation}
where
\(
 \rho_n({\bf r})=\langle \Phi_n |\hat{n}_{\bf r}|\Phi_n\rangle
\)
denotes the particle density
in terms of the number operator $\hat{n}_{\bf r}$.
In any case, 
the $N$ identical particles
must evolve simultaneously as
\(
 \Psi
%({\bf r}_1,{\bf r}_2,\cdots, {\bf r}_N)
\rightarrow 
P_n\Psi/{\sqrt{w_n}},
\)
as they are not strictly independent from each other.
%In general, 
%the $P_n$ for many particle systems
%cannot be expressed as simple sums of
%the single particle counterparts $P_n({\bf r}_i)$.
%
%we cannot express the effect of collapse
%by a one-particle operator.

%So far, we assumed that
%the particle number is constant.
%We have restricted our discussion
%within the non-relativistic limit.
%To go beyond the limits of non-relativistic theory, 

To treat the case where the particle number $N$ is variable,
we have to generalize the collapse prescription furthermore
to realize the localization on the number basis as well.
This is not difficult in principle,
but goes beyond the scope of the paper.

%\bibliography{clps}

\begin{thebibliography}{74}
\expandafter\ifx\csname natexlab\endcsname\relax\def\natexlab#1{#1}\fi
\expandafter\ifx\csname bibnamefont\endcsname\relax
  \def\bibnamefont#1{#1}\fi
\expandafter\ifx\csname bibfnamefont\endcsname\relax
  \def\bibfnamefont#1{#1}\fi
\expandafter\ifx\csname citenamefont\endcsname\relax
  \def\citenamefont#1{#1}\fi
\expandafter\ifx\csname url\endcsname\relax
  \def\url#1{\texttt{#1}}\fi
\expandafter\ifx\csname urlprefix\endcsname\relax\def\urlprefix{URL }\fi
\providecommand{\bibinfo}[2]{#2}
\providecommand{\eprint}[2][]{\url{#2}}

\bibitem[{\citenamefont{Cohen-Tannoudji and Dalibard}(1986)}]{cd86}
\bibinfo{author}{\bibfnamefont{C.}~\bibnamefont{Cohen-Tannoudji}}
  \bibnamefont{and} \bibinfo{author}{\bibfnamefont{J.}~\bibnamefont{Dalibard}},
  \bibinfo{journal}{Europhys. Lett.} \textbf{\bibinfo{volume}{1}},
  \bibinfo{pages}{441} (\bibinfo{year}{1986}).

\bibitem[{\citenamefont{Poratti and Putterman}(1987)}]{pp87}
\bibinfo{author}{\bibfnamefont{M.}~\bibnamefont{Poratti}} \bibnamefont{and}
  \bibinfo{author}{\bibfnamefont{S.}~\bibnamefont{Putterman}},
  \bibinfo{journal}{Phys. Rev.} \textbf{\bibinfo{volume}{A 36}},
  \bibinfo{pages}{929} (\bibinfo{year}{1987});
%\bibitem[{\citenamefont{Poratti and Putterman}(1989)}]{pp89}
\bibinfo{author}{\bibfnamefont{M.}~\bibnamefont{Poratti}} \bibnamefont{and}
  \bibinfo{author}{\bibfnamefont{S.}~\bibnamefont{Putterman}},
  \bibinfo{journal}{Phys. Rev.} \textbf{\bibinfo{volume}{A 39}},
  \bibinfo{pages}{3010} (\bibinfo{year}{1989}).



\bibitem[{\citenamefont{Sauter et~al.}(1986)\citenamefont{Sauter, Blatt,
  Neuhauser, and Toschek}}]{sbnt86}
\bibinfo{author}{\bibfnamefont{T.}~\bibnamefont{Sauter}},
  \bibinfo{author}{\bibfnamefont{R.}~\bibnamefont{Blatt}},
  \bibinfo{author}{\bibfnamefont{W.}~\bibnamefont{Neuhauser}},
  \bibnamefont{and} \bibinfo{author}{\bibfnamefont{P.~E.}
  \bibnamefont{Toschek}}, \bibinfo{journal}{Opt. Commun.}
  \textbf{\bibinfo{volume}{60}}, \bibinfo{pages}{287} (\bibinfo{year}{1986}).

\bibitem[{\citenamefont{Cook}(1988)}]{co88}
\bibinfo{author}{\bibfnamefont{R.}~\bibnamefont{Cook}},
  \bibinfo{journal}{Phys.\ Scr.} \textbf{\bibinfo{volume}{T21}},
  \bibinfo{pages}{49} (\bibinfo{year}{1988}).

\bibitem[{\citenamefont{Blatt and Zoller}(1988)}]{bz88}
\bibinfo{author}{\bibfnamefont{R.}~\bibnamefont{Blatt}} \bibnamefont{and}
  \bibinfo{author}{\bibfnamefont{P.}~\bibnamefont{Zoller}},
  \bibinfo{journal}{Eur. J. Phys.} \textbf{\bibinfo{volume}{9}},
  \bibinfo{pages}{250} (\bibinfo{year}{1988}).

\bibitem[{\citenamefont{Pegg and Knight}(1988)}]{pk88}
\bibinfo{author}{\bibfnamefont{D.~T.} \bibnamefont{Pegg}} \bibnamefont{and}
  \bibinfo{author}{\bibfnamefont{P.~L.} \bibnamefont{Knight}},
  \textbf{\bibinfo{volume}{A37}}, \bibinfo{pages}{4303} (\bibinfo{year}{1988}).

\bibitem[{\citenamefont{Ambrose and Moerner}(1991)}]{am91}
\bibinfo{author}{\bibfnamefont{W.~P.} \bibnamefont{Ambrose}} \bibnamefont{and}
  \bibinfo{author}{\bibfnamefont{W.~E.} \bibnamefont{Moerner}},
  \bibinfo{journal}{Nature} \textbf{\bibinfo{volume}{349}},
  \bibinfo{pages}{225} (\bibinfo{year}{1991}).

\bibitem[{\citenamefont{Ralls and Buhrman}(1988)}]{rb88}
\bibinfo{author}{\bibfnamefont{K.~S.} \bibnamefont{Ralls}} \bibnamefont{and}
  \bibinfo{author}{\bibfnamefont{R.~A.} \bibnamefont{Buhrman}},
  \bibinfo{journal}{Phys. Rev. Lett.} \textbf{\bibinfo{volume}{60}},
  \bibinfo{pages}{2434} (\bibinfo{year}{1988}).

\bibitem[{\citenamefont{Bergquist et~al.}(1986)\citenamefont{Bergquist, Hulet,
  Itano, and Wineland}}]{bhiw86}
\bibinfo{author}{\bibfnamefont{J.~C.} \bibnamefont{Bergquist}},
  \bibinfo{author}{\bibfnamefont{R.~G.} \bibnamefont{Hulet}},
  \bibinfo{author}{\bibfnamefont{W.~M.} \bibnamefont{Itano}}, \bibnamefont{and}
  \bibinfo{author}{\bibfnamefont{D.~J.} \bibnamefont{Wineland}},
  \bibinfo{journal}{Phys. Rev. Lett.} \textbf{\bibinfo{volume}{57}},
  \bibinfo{pages}{1699} (\bibinfo{year}{1986}).

\bibitem[{\citenamefont{Nagourney et~al.}(1986)\citenamefont{Nagourney,
  Sandberg, and Dehmelt}}]{nsd86}
\bibinfo{author}{\bibfnamefont{W.}~\bibnamefont{Nagourney}},
  \bibinfo{author}{\bibfnamefont{J.}~\bibnamefont{Sandberg}}, \bibnamefont{and}
  \bibinfo{author}{\bibfnamefont{H.}~\bibnamefont{Dehmelt}},
  \bibinfo{journal}{Phys. Rev. Lett.} \textbf{\bibinfo{volume}{56}},
  \bibinfo{pages}{2797} (\bibinfo{year}{1986}).

\bibitem[{\citenamefont{Leibfried et~al.}(2003)\citenamefont{Leibfried, Blatt,
  Monroe, and Wineland}}]{lbmw03}
\bibinfo{author}{\bibfnamefont{D.}~\bibnamefont{Leibfried}},
  \bibinfo{author}{\bibfnamefont{R.}~\bibnamefont{Blatt}},
  \bibinfo{author}{\bibfnamefont{C.}~\bibnamefont{Monroe}}, \bibnamefont{and}
  \bibinfo{author}{\bibfnamefont{D.}~\bibnamefont{Wineland}},
  \bibinfo{journal}{Rev. Mod. Phys.} \textbf{\bibinfo{volume}{75}},
  \bibinfo{pages}{281} (\bibinfo{year}{2003}).

\bibitem[{\citenamefont{Bohm}(1952{\natexlab{a}})}]{bo52a}
\bibinfo{author}{\bibfnamefont{D.}~\bibnamefont{Bohm}}, \bibinfo{journal}{Phys.
  Rev.} \textbf{\bibinfo{volume}{85}}, \bibinfo{pages}{166}
  (\bibinfo{year}{1952}{\natexlab{a}});
%\bibitem[{\citenamefont{Bohm}(1952{\natexlab{b}})}]{bo52b}
\bibinfo{author}{\bibfnamefont{D.}~\bibnamefont{Bohm}}, \bibinfo{journal}{Phys.
  Rev.} \textbf{\bibinfo{volume}{85}}, \bibinfo{pages}{180}
  (\bibinfo{year}{1952}{\natexlab{b}});
%\bibitem[{\citenamefont{Bohm and Hiley}(1989)}]{bh}
\bibinfo{author}{\bibfnamefont{D.}~\bibnamefont{Bohm}} \bibnamefont{and}
  \bibinfo{author}{\bibfnamefont{B.}~\bibnamefont{Hiley}},
  \bibinfo{journal}{Phys. Rep.} \textbf{\bibinfo{volume}{172}},
  \bibinfo{pages}{93} (\bibinfo{year}{1989}).

\bibitem[{\citenamefont{Holland}(1993)}]{hol}
\bibinfo{author}{\bibfnamefont{P.}~\bibnamefont{Holland}},
  \emph{\bibinfo{title}{The Quantum Theory of Motion}}
  (\bibinfo{publisher}{Cambridge University Press}, \bibinfo{year}{1993}).

\bibitem[{\citenamefont{D\"urr et~al.}(1992)\citenamefont{D\"urr, Goldstein,
  and Zangh\`{\i}}}]{dgz}
\bibinfo{author}{\bibfnamefont{D.}~\bibnamefont{D\"urr}},
  \bibinfo{author}{\bibfnamefont{S.}~\bibnamefont{Goldstein}},
  \bibnamefont{and}
  \bibinfo{author}{\bibfnamefont{N.}~\bibnamefont{Zangh\`{\i}}},
  \bibinfo{journal}{J. Stat. Phys.} \textbf{\bibinfo{volume}{67}},
  \bibinfo{pages}{843} (\bibinfo{year}{1992}).

\bibitem[{\citenamefont{Ballentine}(1970)}]{b70}
\bibinfo{author}{\bibfnamefont{L.~E.} \bibnamefont{Ballentine}},
  \bibinfo{journal}{Rev. Mod. Phys.} \textbf{\bibinfo{volume}{42}},
  \bibinfo{pages}{358} (\bibinfo{year}{1970}).

\bibitem[{\citenamefont{Zeh}(1970)}]{ze70}
\bibinfo{author}{\bibfnamefont{H.}~\bibnamefont{Zeh}}, \bibinfo{journal}{Found.
  Phys.} \textbf{\bibinfo{volume}{1}}, \bibinfo{pages}{77}
  (\bibinfo{year}{1970}).

\bibitem[{\citenamefont{Machida and Namiki}(1980{\natexlab{a}})}]{mn80a}
\bibinfo{author}{\bibfnamefont{S.}~\bibnamefont{Machida}} \bibnamefont{and}
  \bibinfo{author}{\bibfnamefont{M.}~\bibnamefont{Namiki}},
  \bibinfo{journal}{Prog. Theor. Phys.} \textbf{\bibinfo{volume}{63}},
  \bibinfo{pages}{1457} (\bibinfo{year}{1980}{\natexlab{a}});
%\bibitem[{\citenamefont{Machida and Namiki}(1980{\natexlab{b}})}]{mn80b}
\bibinfo{author}{\bibfnamefont{S.}~\bibnamefont{Machida}} \bibnamefont{and}
  \bibinfo{author}{\bibfnamefont{M.}~\bibnamefont{Namiki}},
  \bibinfo{journal}{Prog. Theor. Phys.} \textbf{\bibinfo{volume}{63}},
  \bibinfo{pages}{1833} (\bibinfo{year}{1980}{\natexlab{b}}).

\bibitem[{\citenamefont{Joos and Zeh}(1985)}]{jz85}
\bibinfo{author}{\bibfnamefont{E.}~\bibnamefont{Joos}} \bibnamefont{and}
  \bibinfo{author}{\bibfnamefont{H.}~\bibnamefont{Zeh}},
  \bibinfo{journal}{Zeit. Phys.} \textbf{\bibinfo{volume}{B 59}},
  \bibinfo{pages}{223} (\bibinfo{year}{1985}).

\bibitem[{\citenamefont{Namiki and Pascazio}(1993)}]{np93}
\bibinfo{author}{\bibfnamefont{M.}~\bibnamefont{Namiki}} \bibnamefont{and}
  \bibinfo{author}{\bibfnamefont{S.}~\bibnamefont{Pascazio}},
  \bibinfo{journal}{Phys. Rep.} \textbf{\bibinfo{volume}{232}},
  \bibinfo{pages}{301} (\bibinfo{year}{1993}).

\bibitem[{\citenamefont{Giulini et~al.}(1996)\citenamefont{Giulini, Joos,
  Kiefer, Kumpsch, Stamatescu, and Zeh}}]{gjkksz96}
\bibinfo{author}{\bibfnamefont{D.}~\bibnamefont{Giulini}},
  \bibinfo{author}{\bibfnamefont{E.}~\bibnamefont{Joos}},
  \bibinfo{author}{\bibfnamefont{C.}~\bibnamefont{Kiefer}},
  \bibinfo{author}{\bibfnamefont{J.}~\bibnamefont{Kumpsch}},
  \bibinfo{author}{\bibfnamefont{I.-O.} \bibnamefont{Stamatescu}},
  \bibnamefont{and} \bibinfo{author}{\bibfnamefont{H.}~\bibnamefont{Zeh}},
  \emph{\bibinfo{title}{Decoherence and the Appearance of a Classical World in
  Quantum Theory}} (\bibinfo{publisher}{Springer-Verlag},
  \bibinfo{year}{1996}).

\bibitem[{\citenamefont{Zurek}(2003)}]{zu03}
\bibinfo{author}{\bibfnamefont{W.}~\bibnamefont{Zurek}}, \bibinfo{journal}{Rev.
  Mod. Phys.} \textbf{\bibinfo{volume}{75}}, \bibinfo{pages}{715}
  (\bibinfo{year}{2003}).

\bibitem[{\citenamefont{van Frassen}(1991)}]{vF91}
\bibinfo{author}{\bibfnamefont{B.}~\bibnamefont{van Frassen}},
  \emph{\bibinfo{title}{Quantum Mechanics: an Empiricist View}}
  (\bibinfo{publisher}{Clarendon Press, Oxford}, \bibinfo{year}{1991}).

\bibitem[{\citenamefont{Healey}(1989)}]{he89}
\bibinfo{author}{\bibfnamefont{R.}~\bibnamefont{Healey}},
  \emph{\bibinfo{title}{The Philosophy of Quantum Mechanics: an Interactive
  Interpretation}} (\bibinfo{publisher}{Cambridge University Press},
  \bibinfo{year}{1989}).

\bibitem[{\citenamefont{Vermaas and Dieks}(1998)}]{vd98}
\bibinfo{author}{\bibfnamefont{P.}~\bibnamefont{Vermaas}} \bibnamefont{and}
  \bibinfo{author}{\bibfnamefont{D.}~\bibnamefont{Dieks}},
  \emph{\bibinfo{title}{The Modal Interpretation of Quantum Mechanics}}
  (\bibinfo{publisher}{Kluwer Academic Publishers, Dordrecht},
  \bibinfo{year}{1998}).

\bibitem[{\citenamefont{Griffiths}(1984)}]{gr84}
\bibinfo{author}{\bibfnamefont{R.}~\bibnamefont{Griffiths}},
  \bibinfo{journal}{J. Stat. Phys.} \textbf{\bibinfo{volume}{36}},
  \bibinfo{pages}{219} (\bibinfo{year}{1984});
%\bibitem[{\citenamefont{Griffiths}(1996)}]{gr96}
\bibinfo{author}{\bibfnamefont{R.}~\bibnamefont{Griffiths}},
  \bibinfo{journal}{Phys. Rev.} \textbf{\bibinfo{volume}{A 54}},
  \bibinfo{pages}{2759} (\bibinfo{year}{1996}).

\bibitem[{\citenamefont{Omn\`es}(1992)}]{om92}
\bibinfo{author}{\bibfnamefont{R.}~\bibnamefont{Omn\`es}},
  \bibinfo{journal}{Rev. Mod. Phys.} \textbf{\bibinfo{volume}{64}},
  \bibinfo{pages}{339} (\bibinfo{year}{1992}).

\bibitem[{\citenamefont{Gell-Mann and Hartle}(1990)}]{gh90}
\bibinfo{author}{\bibfnamefont{M.}~\bibnamefont{Gell-Mann}} \bibnamefont{and}
  \bibinfo{author}{\bibfnamefont{J.}~\bibnamefont{Hartle}}, in
  \emph{\bibinfo{booktitle}{Complexity, Entropy, and the Physics of
  Information}}, edited by \bibinfo{editor}{\bibnamefont{W.H.Zurek}}
  (\bibinfo{publisher}{Addison Wesley Publishing Company, Reading},
  \bibinfo{year}{1990}).

\bibitem[{\citenamefont{{Everett III}}(1957)}]{ev57}
\bibinfo{author}{\bibfnamefont{H.}~\bibnamefont{{Everett III}}},
  \bibinfo{journal}{Rev. Mod. Phys.} \textbf{\bibinfo{volume}{29}},
  \bibinfo{pages}{454} (\bibinfo{year}{1957}).

\bibitem[{\citenamefont{DeWitt and Graham}(1973)}]{dg73}
\bibinfo{author}{\bibfnamefont{D.}~\bibnamefont{DeWitt}} \bibnamefont{and}
  \bibinfo{author}{\bibfnamefont{N.}~\bibnamefont{Graham}},
  \emph{\bibinfo{title}{The Many-Worlds Interpretation of Quantum Mechanics}}
  (\bibinfo{publisher}{Princeton University Press}, \bibinfo{year}{1973}).

\bibitem[{\citenamefont{Deutsch}(1985)}]{de85}
\bibinfo{author}{\bibfnamefont{D.}~\bibnamefont{Deutsch}},
  \bibinfo{journal}{Int. Journ. Theor. Phys.} \textbf{\bibinfo{volume}{24}},
  \bibinfo{pages}{1} (\bibinfo{year}{1985}).

\bibitem[{\citenamefont{Albert and Loewer}(1988)}]{al}
\bibinfo{author}{\bibfnamefont{D.}~\bibnamefont{Albert}} \bibnamefont{and}
  \bibinfo{author}{\bibfnamefont{B.}~\bibnamefont{Loewer}},
  \bibinfo{journal}{Synthese} \textbf{\bibinfo{volume}{77}},
  \bibinfo{pages}{195} (\bibinfo{year}{1988}).

\bibitem[{\citenamefont{Albert}(1992)}]{a92}
\bibinfo{author}{\bibfnamefont{D.}~\bibnamefont{Albert}},
  \emph{\bibinfo{title}{Quantum Mechanics and Experience}}
  (\bibinfo{publisher}{Harward University Press, Cambridge, MA},
  \bibinfo{year}{1992}).

\bibitem[{\citenamefont{Bell}(1981)}]{bell87qmc}
\bibinfo{author}{\bibfnamefont{J.}~\bibnamefont{Bell}}, in
  \emph{\bibinfo{booktitle}{Quantum Gravity 2}}, edited by
  \bibinfo{editor}{\bibfnamefont{C.}~\bibnamefont{Isham}},
  \bibinfo{editor}{\bibfnamefont{R.}~\bibnamefont{Pensrose}}, \bibnamefont{and}
  \bibinfo{editor}{\bibfnamefont{D.}~\bibnamefont{Sciama}}
  (\bibinfo{year}{1981}), \bibinfo{note}{also in: {\it Speakable and
  unspeakable in quantum mechanics}, Cambridge University Press, 117 (1987)}.

\bibitem[{\citenamefont{Bell}(1990)}]{be90}
\bibinfo{author}{\bibfnamefont{J.}~\bibnamefont{Bell}}, in
  \emph{\bibinfo{booktitle}{Sixty-Two Years of Uncertainity}}, edited by
  \bibinfo{editor}{\bibfnamefont{A.}~\bibnamefont{Miller}}
  (\bibinfo{publisher}{Plenum Press, New York}, \bibinfo{year}{1990}),
  \bibinfo{note}{also in, Physics World {\bf 3}, 33 (1990).}

\bibitem[{\citenamefont{Bohm and Bub}(1966)}]{bb66}
\bibinfo{author}{\bibfnamefont{D.}~\bibnamefont{Bohm}} \bibnamefont{and}
  \bibinfo{author}{\bibfnamefont{J.}~\bibnamefont{Bub}}, \bibinfo{journal}{Rev.
  Mod. Phys.} \textbf{\bibinfo{volume}{38}}, \bibinfo{pages}{453}
  (\bibinfo{year}{1966}).

\bibitem[{\citenamefont{Pearle}(1976)}]{pe76}
\bibinfo{author}{\bibfnamefont{P.}~\bibnamefont{Pearle}},
  \bibinfo{journal}{Phys. Rev.} \textbf{\bibinfo{volume}{D 13}},
  \bibinfo{pages}{857} (\bibinfo{year}{1976});
%\bibitem[{\citenamefont{Pearle}(1984)}]{pe84}
\bibinfo{author}{\bibfnamefont{P.}~\bibnamefont{Pearle}},
  \bibinfo{journal}{Phys. Rev. Lett.} \textbf{\bibinfo{volume}{53}},
  \bibinfo{pages}{1775} (\bibinfo{year}{1984});
%\bibitem[{\citenamefont{Pearle}(1989)}]{pe89}
\bibinfo{author}{\bibfnamefont{P.}~\bibnamefont{Pearle}},
  \bibinfo{journal}{Phys. Rev.} \textbf{\bibinfo{volume}{A 39}},
  \bibinfo{pages}{2277} (\bibinfo{year}{1989}).

\bibitem[{\citenamefont{Di\'osi}(1985)}]{di85}
\bibinfo{author}{\bibfnamefont{L.}~\bibnamefont{Di\'osi}},
  \bibinfo{journal}{Phys. Lett.} \textbf{\bibinfo{volume}{A 112}},
  \bibinfo{pages}{288} (\bibinfo{year}{1985});
%\bibitem[{\citenamefont{Di\'osi}(1987)}]{di87}
\bibinfo{author}{\bibfnamefont{L.}~\bibnamefont{Di\'osi}},
  \bibinfo{journal}{Phys. Lett.} \textbf{\bibinfo{volume}{A 122}},
  \bibinfo{pages}{221} (\bibinfo{year}{1987});
%\bibitem[{\citenamefont{Di\'osi}(1988)}]{di88}
\bibinfo{author}{\bibfnamefont{L.}~\bibnamefont{Di\'osi}},
  \bibinfo{journal}{Phys. Lett.} \textbf{\bibinfo{volume}{A 129}},
  \bibinfo{pages}{419} (\bibinfo{year}{1988}).

\bibitem[{\citenamefont{Gisin}(1984{\natexlab{a}})}]{gi84a}
\bibinfo{author}{\bibfnamefont{N.}~\bibnamefont{Gisin}},
  \bibinfo{journal}{Phys. Rev. Lett.} \textbf{\bibinfo{volume}{52}},
  \bibinfo{pages}{1657} (\bibinfo{year}{1984}{\natexlab{a}});
%\bibitem[{\citenamefont{Gisin}(1984{\natexlab{b}})}]{gi84b}
\bibinfo{author}{\bibfnamefont{N.}~\bibnamefont{Gisin}},
  \bibinfo{journal}{Phys. Rev. Lett.} \textbf{\bibinfo{volume}{53}},
  \bibinfo{pages}{1776} (\bibinfo{year}{1984}{\natexlab{b}}).

\bibitem[{\citenamefont{Ghirardi et~al.}(1986)\citenamefont{Ghirardi, Rimini,
  and Weber}}]{grw86}
\bibinfo{author}{\bibfnamefont{G.~C.} \bibnamefont{Ghirardi}},
  \bibinfo{author}{\bibfnamefont{A.}~\bibnamefont{Rimini}}, \bibnamefont{and}
  \bibinfo{author}{\bibfnamefont{T.}~\bibnamefont{Weber}},
  \bibinfo{journal}{Phys.\ Rev.} \textbf{\bibinfo{volume}{D34}},
  \bibinfo{pages}{470} (\bibinfo{year}{1986}).

\bibitem[{\citenamefont{Bell}(1987{\natexlab{a}})}]{bell87qj}
\bibinfo{author}{\bibfnamefont{J.}~\bibnamefont{Bell}}, in
  \emph{\bibinfo{booktitle}{Schr\"odinger: Centanary Celebration of a
  Polymath}}, edited by
  \bibinfo{editor}{\bibfnamefont{C.}~\bibnamefont{Kilmister}}
  (\bibinfo{publisher}{Cambridge University Press, Cambridge},
  \bibinfo{year}{1987}{\natexlab{a}}), \bibinfo{note}{also in, {\it Speakable
  and unspeakable in quantum mechanics}, Cambridge University Press, 201
  (1987)}.

\bibitem[{\citenamefont{Ghirardi et~al.}(1990)\citenamefont{Ghirardi, Pearle,
  and Rimini}}]{gpr90}
\bibinfo{author}{\bibfnamefont{G.}~\bibnamefont{Ghirardi}},
  \bibinfo{author}{\bibfnamefont{P.}~\bibnamefont{Pearle}}, \bibnamefont{and}
  \bibinfo{author}{\bibfnamefont{A.}~\bibnamefont{Rimini}},
  \bibinfo{journal}{Phys. Rev.} \textbf{\bibinfo{volume}{A 42}},
  \bibinfo{pages}{78} (\bibinfo{year}{1990}).

\bibitem[{\citenamefont{Ghirardi}(2000)}]{gh00}
\bibinfo{author}{\bibfnamefont{G.}~\bibnamefont{Ghirardi}},
  \bibinfo{journal}{Found. Phys.} \textbf{\bibinfo{volume}{30}},
  \bibinfo{pages}{1337} (\bibinfo{year}{2000}).

\bibitem[{\citenamefont{Bassi and Ghirardi}(2003)}]{bg03}
\bibinfo{author}{\bibfnamefont{A.}~\bibnamefont{Bassi}} \bibnamefont{and}
  \bibinfo{author}{\bibfnamefont{G.}~\bibnamefont{Ghirardi}},
  \bibinfo{journal}{Phys. Rep.} \textbf{\bibinfo{volume}{379}},
  \bibinfo{pages}{257} (\bibinfo{year}{2003}).

\bibitem[{\citenamefont{Pearle}(1999)}]{p99}
\bibinfo{author}{\bibfnamefont{P.}~\bibnamefont{Pearle}}, in
  \emph{\bibinfo{booktitle}{Open Systems and Measurement in Relativistic
  Quantum Theory}}, edited by \bibinfo{editor}{\bibnamefont{F.Petruccione}}
  \bibnamefont{and} \bibinfo{editor}{\bibnamefont{H.P.Breuer}}
  (\bibinfo{publisher}{Springer Verlag}, \bibinfo{year}{1999}).

\bibitem[{\citenamefont{Pearle et~al.}(1999)\citenamefont{Pearle, Ring, Collar,
  and {Avignone III}}}]{prca99}
\bibinfo{author}{\bibfnamefont{P.}~\bibnamefont{Pearle}},
  \bibinfo{author}{\bibfnamefont{J.}~\bibnamefont{Ring}},
  \bibinfo{author}{\bibfnamefont{J.~I.} \bibnamefont{Collar}},
  \bibnamefont{and} \bibinfo{author}{\bibfnamefont{F.~T.}
  \bibnamefont{{Avignone III}}}, \bibinfo{journal}{Found. Phys.}
  \textbf{\bibinfo{volume}{29}}, \bibinfo{pages}{465} (\bibinfo{year}{1999}).

\bibitem[{\citenamefont{Frenkel}(1990)}]{fr90}
\bibinfo{author}{\bibfnamefont{A.}~\bibnamefont{Frenkel}},
  \bibinfo{journal}{Found. Phys.} \textbf{\bibinfo{volume}{20}},
  \bibinfo{pages}{159} (\bibinfo{year}{1990}).

\bibitem[{\citenamefont{Milburn}(1991)}]{mi91}
\bibinfo{author}{\bibfnamefont{G.~J.} \bibnamefont{Milburn}},
  \bibinfo{journal}{Phys. Rev.} \textbf{\bibinfo{volume}{A44}},
  \bibinfo{pages}{5401} (\bibinfo{year}{1991}).

\bibitem[{\citenamefont{Percival}(1995)}]{pe95}
\bibinfo{author}{\bibfnamefont{I.~C.} \bibnamefont{Percival}},
  \bibinfo{journal}{Proc. R. Soc. Lond.} \textbf{\bibinfo{volume}{A451}},
  \bibinfo{pages}{503} (\bibinfo{year}{1995}).

\bibitem[{\citenamefont{Hughston}(1996)}]{hu96}
\bibinfo{author}{\bibfnamefont{L.}~\bibnamefont{Hughston}},
  \bibinfo{journal}{Proc. R. Soc. London} \textbf{\bibinfo{volume}{A 452}},
  \bibinfo{pages}{953} (\bibinfo{year}{1996}).

\bibitem[{\citenamefont{Adler and Howritz}(2000)}]{ah00}
\bibinfo{author}{\bibfnamefont{S.}~\bibnamefont{Adler}} \bibnamefont{and}
  \bibinfo{author}{\bibfnamefont{P.}~\bibnamefont{Howritz}},
  \bibinfo{journal}{J. Math. Phys.} \textbf{\bibinfo{volume}{41}},
  \bibinfo{pages}{2485} (\bibinfo{year}{2000});
%\bibitem[{\citenamefont{Adler and Brun}(2001)}]{ab01}
\bibinfo{author}{\bibfnamefont{S.}~\bibnamefont{Adler}} \bibnamefont{and}
  \bibinfo{author}{\bibfnamefont{T.}~\bibnamefont{Brun}}, \bibinfo{journal}{J.
  Phys.} \textbf{\bibinfo{volume}{A 34}}, \bibinfo{pages}{4797}
  (\bibinfo{year}{2001});
%\bibitem[{\citenamefont{Adler}(2002)}]{ad02}
\bibinfo{author}{\bibfnamefont{S.}~\bibnamefont{Adler}}, \bibinfo{journal}{J.
  Phys.} \textbf{\bibinfo{volume}{A 35}}, \bibinfo{pages}{841}
  (\bibinfo{year}{2002}).

\bibitem[{toa()}]{toappear}
\bibinfo{note}{A preliminary version of the work is to appear in Opt. Spectr.;
  T. Okabe, quant-ph/0410095}.

\bibitem[{\citenamefont{Tegmark}(1993)}]{te93}
\bibinfo{author}{\bibfnamefont{M.}~\bibnamefont{Tegmark}},
  \bibinfo{journal}{Found. Phys. Lett.} \textbf{\bibinfo{volume}{6}},
  \bibinfo{pages}{571} (\bibinfo{year}{1993}).

\bibitem[{\citenamefont{Tessieri et~al.}(1995)\citenamefont{Tessieri, Vitali,
  and Grigolini}}]{tvg95}
\bibinfo{author}{\bibfnamefont{L.}~\bibnamefont{Tessieri}},
  \bibinfo{author}{\bibfnamefont{D.}~\bibnamefont{Vitali}}, \bibnamefont{and}
  \bibinfo{author}{\bibfnamefont{P.}~\bibnamefont{Grigolini}},
  \bibinfo{journal}{Phys. Rev.} \textbf{\bibinfo{volume}{A51}},
  \bibinfo{pages}{4404} (\bibinfo{year}{1995}).

\bibitem[{\citenamefont{Misra and Sudarshan}(1977)}]{ms77}
\bibinfo{author}{\bibfnamefont{B.}~\bibnamefont{Misra}} \bibnamefont{and}
  \bibinfo{author}{\bibfnamefont{E.}~\bibnamefont{Sudarshan}},
  \bibinfo{journal}{J. Math. Phys.} \textbf{\bibinfo{volume}{18}},
  \bibinfo{pages}{756} (\bibinfo{year}{1977}).

\bibitem[{\citenamefont{Peres}(1980)}]{pe80}
\bibinfo{author}{\bibfnamefont{A.}~\bibnamefont{Peres}}, \bibinfo{journal}{Am.
  J. Phys.} \textbf{\bibinfo{volume}{48}}, \bibinfo{pages}{931}
  (\bibinfo{year}{1980}).

\bibitem[{\citenamefont{Kraus}(1980)}]{kr80}
\bibinfo{author}{\bibfnamefont{K.}~\bibnamefont{Kraus}},
  \bibinfo{journal}{Found. Phys.} \textbf{\bibinfo{volume}{11}},
  \bibinfo{pages}{547} (\bibinfo{year}{1980}).

\bibitem[{\citenamefont{Gallis and Fleming}(1990)}]{gf90}
\bibinfo{author}{\bibfnamefont{M.~R.} \bibnamefont{Gallis}} \bibnamefont{and}
  \bibinfo{author}{\bibfnamefont{G.~N.} \bibnamefont{Fleming}},
  \bibinfo{journal}{Phys. Rev.} \textbf{\bibinfo{volume}{A42}},
  \bibinfo{pages}{38} (\bibinfo{year}{1990}).

\bibitem[{\citenamefont{Aharonov and Albert}(1981)}]{aa81}
\bibinfo{author}{\bibfnamefont{Y.}~\bibnamefont{Aharonov}} \bibnamefont{and}
  \bibinfo{author}{\bibfnamefont{D.}~\bibnamefont{Albert}},
  \bibinfo{journal}{Phys. Rev.} \textbf{\bibinfo{volume}{D 24}},
  \bibinfo{pages}{359} (\bibinfo{year}{1981});
%\bibitem[{\citenamefont{Aharonov and Albert}(1984)}]{aa84}
\bibinfo{author}{\bibfnamefont{Y.}~\bibnamefont{Aharonov}} \bibnamefont{and}
  \bibinfo{author}{\bibfnamefont{D.}~\bibnamefont{Albert}},
  \bibinfo{journal}{Phys. Rev.} \textbf{\bibinfo{volume}{D 29}},
  \bibinfo{pages}{228} (\bibinfo{year}{1984}).

\bibitem[{\citenamefont{Eberhard}(1978)}]{eb78}
\bibinfo{author}{\bibfnamefont{P.}~\bibnamefont{Eberhard}},
  \bibinfo{journal}{Nuovo Cimento} \textbf{\bibinfo{volume}{B 46}},
  \bibinfo{pages}{392} (\bibinfo{year}{1978}).

\bibitem[{\citenamefont{Bell}(1987{\natexlab{b}})}]{bell87bqft}
\bibinfo{author}{\bibfnamefont{J.}~\bibnamefont{Bell}}, in
  \emph{\bibinfo{booktitle}{Speakable and unspeakable in quantum mechanics}}
  (\bibinfo{publisher}{Cambridge University Press},
  \bibinfo{year}{1987}{\natexlab{b}}), p. \bibinfo{pages}{173}.

\bibitem[{not()}]{noteadded}
\bibinfo{note}{Then the special frame would be experimentally distinguishable
  by a dilation of the collapse time $\tau_0$.}

\end{thebibliography}

\end{document}